\documentclass[preprint2]{aastex}

\newcommand{\henb}[2]{\frac{\partial #1}{\partial #2}}

\slugcomment{}

\shorttitle{Parametric Study of the RWI}
\shortauthors{Ono et al.}

\begin{document}


\title{Parametric Study of the Rossby Wave Instability in a Two-Dimensional Barotropic Disk}




\author{Tomohiro Ono}
\affil{Department of Astronomy, Graduate School of Science, Kyoto University, Sakyo-ku, Kyoto 606-8502, Japan}
\email{ono.t@kusastro.kyoto-u.ac.jp}

\and
\author{Takayuki Muto}
\affil{Division of Liberal Arts, Kogakuin University, 1-24-2 Nishi-Shinjuku, Shinjuku-ku, Tokyo 163-8677, Japan}

\and
\author{Taku Takeuchi and Hideko Nomura}
\affil{Department of Earth and Planetary Sciences, Tokyo Institute of Technology, Ookayama, Meguro-ku, Tokyo 152-8551, Japan}




\begin{abstract}
Protoplanetary disks with non-axisymmetric structures have been observed. 
The Rossby wave instability (RWI) is considered as one of the origins of the non-axisymmetric structures. 
We perform linear stability analyses of the RWI in barotropic flow using four representative types of the background flow on a wide parameter space. 
We find that the co-rotation radius is located at the background vortensity minimum with large concavity if the system is marginally stable to the RWI, and this allows us to check the stability against the RWI easily. 
We newly derive the {\it necessary and sufficient} condition for the onset of the RWI in semi-analytic form. 
We discuss the applicability of the new condition in realistic systems and the physical nature of the RWI. 
\end{abstract}





\section{Introduction}

Protoplanetary disks are the birthplace of planets. 
Recent observations have revealed protoplanetary disks having an inner cavity and strongly non-axisymmetric structures in the outer region \citep[e.g.,][]{2013Sci...340.1199V, 2013PASJ...65L..14F}. 
In such protoplanetary disks, dust distribution is much more non-axisymmetric than gas distribution. 
It is often considered that the non-axisymmetric structures as observed are formed by a gas large-scale vortex \citep[e.g.,][]{2014ApJ...783L..13P}, which captures large dust particles ($\sim$~mm in size) due to gas drag. 
When protoplanetary disks have a rapid radial variation, such as boundaries between magnetically active and non-active regions \citep[''dead zones'';][]{1996ApJ...457..355G} or edges of gaps induced by a giant planet \citep{1986ApJ...309..846L}, the gas large-scale vortex is formed \citep[e.g.,][]{1995A&A...295L...1B, 2005ApJ...624.1003L, 2013MNRAS.433.2626R}. 

The gas large-scale vortex can be formed by the Rossby wave instability (RWI), which is a non-axisymmetric hydrodynamic instability. 
The RWI was first studied with two-dimensional (2D) linear stability analyses \citep{1978ApJ...221...51L, 1999ApJ...513..805L, 2000ApJ...533.1023L, 2010A&A...521A..25U}. 
These previous works show that protoplanetary disks with the rapid radial variation, such as a step jump up (as Figure \ref{fig1}) and a bump (as Figure \ref{fig3}) in the surface mass density profile, are unstable to non-axisymmetric perturbations, and the growth rate of the RWI is of the order of dynamical timescale. 
A number of 2D non-linear hydrodynamic simulations are carried out and, in general, the results are consistent with linear stability analyses \citep[e.g.,][]{2001ApJ...551..874L, 2006ApJ...649..415I, 2006A&A...446L..13V, 2008A&A...491L..41L, 2009A&A...497..869L, 2009ApJ...690L..52L, 2010MNRAS.405.1473L, 2011MNRAS.415.1426L}. 
The RWI in three-dimensional (3D) disks has also been studied using both linear stability analyses \citep[e.g.,][]{2012ApJ...754...21L, 2013ApJ...765...84L, 2012MNRAS.422.2399M} and numerical simulations \citep[e.g.,][]{2010A&A...516A..31M, 2012A&A...542A...9M, 2012MNRAS.426.3211L, 2012ApJ...756...62L, 2013A&A...559A..30R}. 
It is indicated that, at least qualitatively, 2D calculations resemble 3D results \citep{2012ApJ...754...21L, 2013A&A...559A..30R}. 
Therefore, essential physics of the RWI may be captured within 2D frameworks. 
\citet{2010A&A...521A..25U} points out that the RWI is similar to the shear instability for incompressible flow in shallow water. 
It is also known that the RWI is, to some extent, similar to the Drury-Papaloizou-Pringle instability \citep{1980MNRAS.193..337D, 1985MNRAS.217..821D, 1984MNRAS.208..721P, 1985MNRAS.213..799P, 1987MNRAS.225..267P, 1986MNRAS.221..339G, 1987MNRAS.228....1N}. 
 
\citet{1999ApJ...513..805L} derives a condition for the RWI (Lovelace's condition), which reads that background flow should have at least one extremum in the vortensity profile. 
However, the sufficiency of the Lovelace's condition against the RWI has not been investigated well. 
\citet{2000ApJ...533.1023L} performs 2D linear stability analyses for both barotropic and non-barotropic flows, and calculates the growth rate for each azimuthal mode number. 
In this paper, we perform linear stability analyses for only barotropic flow with a similar method to \citet{2000ApJ...533.1023L}, but using more various types of background flow on a wider parameter space. 
We derive newly the {\it necessary and sufficient} condition for the onset of the RWI in semi-analytic form, which, we consider, can be used generally.  
We also discuss the physical nature of the RWI from results of the linear stability analyses. 

We show formulation in \S 2 and briefly review the known conditions associated with the RWI in \S 3. 
The stability of the system against the RWI is given in \S 4. 
We investigate marginally stable states to the RWI and derive a new condition for the RWI in \S 5. 
In \S 6, we discuss the physical nature of the RWI.
Conclusions of this work are summarized in \S 7.

\section{Formulation}
\subsection{Basic Equations}
We consider geometrically thin, non-magnetized and inviscid disks. 
This is because the essence of the RWI seems to be understood in the 2D and purely hydrodynamic regime. 
We use a cylindrical coordinate system with ($r, \varphi$) being in the disk plane and the origin at the central star. 
We assume that self-gravity of the disk is negligible. 
The gravitational potential of the central star is denoted by $\Phi (r) = -GM/r$, where $M$ is the mass of the central star, and the disk surface mass density profile is given by $\Sigma(r, \varphi, t)$. 
We denote (vertically integrated) pressure by $P(r, \varphi, t)$ and assume, for simplicity, that the disk gas is barotropic, i.e., $P=P(\Sigma)\propto\Sigma^{\Gamma}$, where $\Gamma$ is the effective adiabatic index of the gas. 
The adiabatic sound speed $c(r, \varphi, t)$ is given by $c = \sqrt{\Gamma P / \Sigma}$. 

The continuity equation is 
\begin{equation}
\henb{\Sigma}{t} + \frac{1}{r}\henb{}{r} (r \Sigma v_r)+\frac{1}{r}\henb{}{\varphi}(\Sigma v_\varphi)=0,
\end{equation}
where $t$ is time, ${\bf v}(r, \varphi, t) \equiv v_r (r, \varphi, t) {\bf \hat{r}} + v_\varphi (r, \varphi, t) {\bf \hat{\varphi}}$ is velocity field, ${\bf \hat{r}}$ is the unit vector in the $r$ direction and ${\bf \hat{\varphi}}$ is the unit vector in the $\varphi$ direction. 
The two components ($r, \varphi$) of the equations of motion are   
\begin{eqnarray}
\henb{v_r}{t} + v_r\henb{v_r}{r}+\frac{v_\varphi}{r}\henb{v_r}{\varphi}-\frac{v_\varphi^2}{r}=-\frac{GM}{r^2}-\frac{1}{\Sigma}\henb{P}{r}, \\
\henb{v_\varphi}{t} + v_r\henb{v_\varphi}{r}+\frac{v_\varphi}{r}\henb{v_\varphi}{\varphi}+\frac{v_r v_\varphi}{r}=-\frac{1}{r \Sigma}\henb{P}{\varphi}. 
\end{eqnarray}

From equations (1)-(3), it is possible to obtain the equation of vortensity conservation with respect to fluid elements; 
\begin{equation}
\henb{q}{t}+v_r \henb{q}{r} +\frac{v_\varphi}{r}\henb{q}{\varphi}=0, 
\end{equation}
where $q(r, \varphi, t)\equiv (\mathrm{rot}\, {\bf v})_z/\Sigma$ is vortensity. 

It should be noted that we have assumed, for simplicity, that the barycenter of the system always coincide with the central star, and is located at the origin. 
This means that the indirect term of the gravitational potential, which arises from the deviation of the barycenter of the system from the position of the central star, is negligible. 
The indirect term can have a significant effect on the disk morphology if the disk is massive \citep[e.g.,][]{2015ApJ...798L..25M, 2015arXiv151103497Z}. 
In this paper, we restrict ourselves to consider low-mass disks.  
We expect that the essential physics of the RWI can be captured within our framework. 

\subsection{Background Profiles}
To perform linear perturbation analyses, we first need to define profiles of the background disks. 
We use the background profiles being stationary ($\partial/\partial t = 0$) and axisymmetric ($\partial/\partial \varphi = 0$), which denoted by subscript ''0'', e.g., $\Sigma_0(r), P_0(r)$ and ${\bf v}_0(r)$. 
Here, ${\bf v}_0(r) = v_{\varphi 0} (r) \bf{\hat{\varphi}}$ is background velocity field and $v_{r0}=0$ is satisfied. 

We assume that the background profiles have radially smooth "base" profiles and rapid radial variations. 
As a "base" profile, the surface mass density is given by a simple power-law profile with the index $-\mathfrak{f}$, $\Sigma_\mathrm{b}(r) = \Sigma_\mathrm{n} (r/r_\mathrm{n})^{-\mathfrak{f}}$, where $r_\mathrm{n}$ is a representative radius and $\Sigma_\mathrm{n}$ is the surface mass density of this base disk model at $r=r_\mathrm{n}$. 
As a form of $\Sigma_0 / \Sigma_\mathrm{b}$, we consider four representative cases, namely, step jump up (SJU), step jump down (SJD), Gaussian bump (GB), and Gaussian gap (GG). 
All of them are represented by two parameters: $\mathcal{A}$ for the amplitude of the variation and $\Delta w$ for the width. 
The specific forms of these four profiles are as follows. 
\begin{itemize}
\item Step Jump Up (SJU)\\
$\Sigma_0 / \Sigma_\mathrm{b}$ is given by 
\begin{equation}
\frac{\Sigma_0}{\Sigma_\mathrm{b}} =1+\frac{\mathcal{A}}{2}\left[ \mathrm{tanh}\left(\frac{r-r_\mathrm{n}}{\Delta w} \right)+1 \right].
\end{equation}
The variation of $\Sigma_0 /\Sigma_\mathrm{b}$ is a smooth jump up from $1$ to ($\mathcal{A}+1$) over $\Delta w$ around $r=r_\mathrm{n}$. 
We call this the step jump up (SJU) type. 
The SJU type is same as the ''HSJ'' type in \citet{2000ApJ...533.1023L}. 
\begin{figure}
\begin{center}
\epsscale{1.0}
\plotone{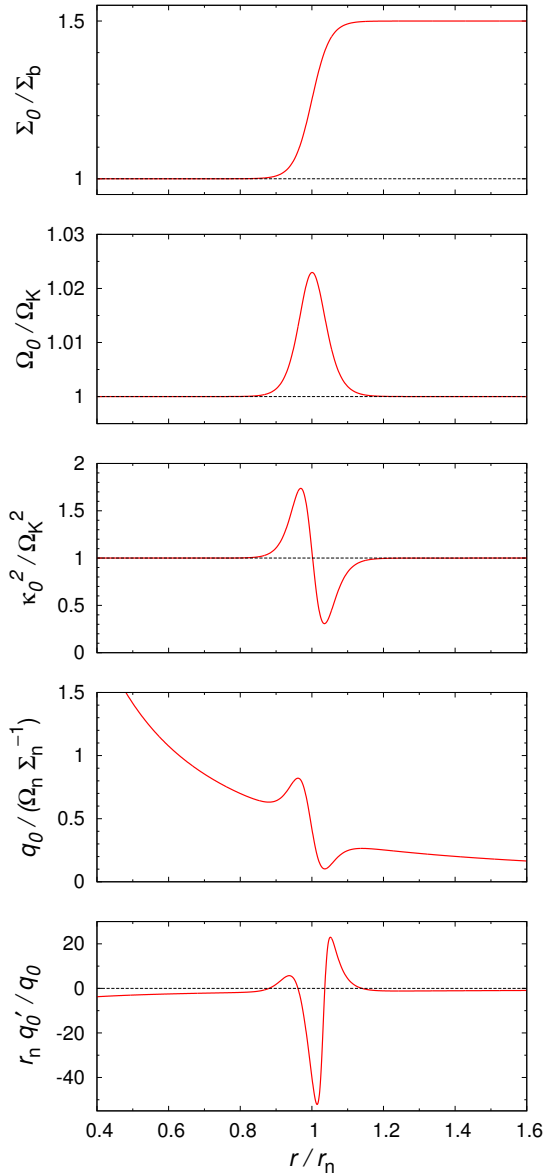}
\caption{$\Sigma_0/\Sigma_\mathrm{b}$, $\Omega_0/\Omega_\mathrm{K}$, $\kappa^2_0/\Omega_\mathrm{K}^2$, $q_0$ and $q_0^\prime /q_0$ are shown for the step jump up (SJU) type with $\mathcal{A}=0.5$, $\Delta w =0.05 r_\mathrm{n}$, $h=0.1$ and $\Gamma=5/3$. 
The variation of $\Sigma_0/\Sigma_\mathrm{b}$ is a smooth jump up from $1$ to $(\mathcal{A}+1)$ over $\Delta w$ around $r=r_\mathrm{n}$. \label{fig1}}
\end{center}
\end{figure}

\item Step Jump Down (SJD)\\
$\Sigma_0 / \Sigma_\mathrm{b}$ is given by 
\begin{equation}
\frac{\Sigma_0}{\Sigma_\mathrm{b}} =1+\frac{\mathcal{A}}{2}\left[ 1-\mathrm{tanh}\left(\frac{r-r_\mathrm{n}}{\Delta w} \right) \right].
\end{equation}
The variation of $\Sigma_0 /\Sigma_\mathrm{b}$ is a smooth jump down from ($\mathcal{A}+1$) to $1$ over $\Delta w$ around $r=r_\mathrm{n}$. 
We call this the step jump down (SJD) type. 
It is noted that the SJD type is not equivalent to the SJU type due to curvature effects even if $\Sigma_\mathrm{b}$ is constant. 
\begin{figure}
\begin{center}
\epsscale{1.0}
\plotone{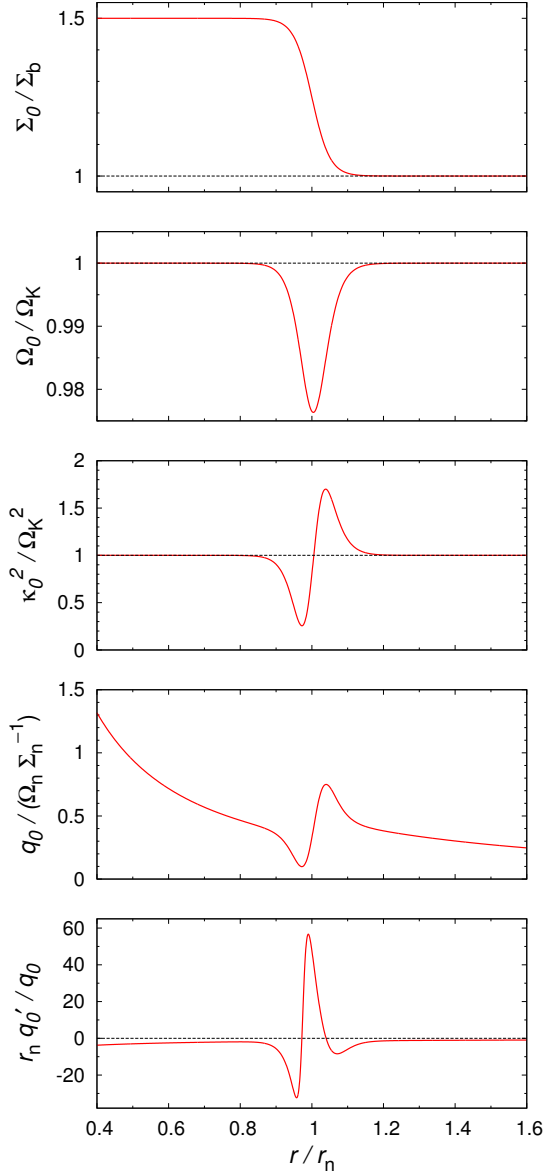}
\caption{Similar to Figure \ref{fig1}, but for the step jump down (SJD) type with $\mathcal{A}=0.5$, $\Delta w =0.05 r_\mathrm{n}$, $h=0.1$ and $\Gamma=5/3$. 
The variation of $\Sigma_0 /\Sigma_\mathrm{b}$ is a smooth jump down from ($\mathcal{A}+1$) to $1$ over $\Delta w$ around $r=r_\mathrm{n}$.  \label{fig2}}
\end{center}
\end{figure}

\item Gaussian Bump (GB)\\
$\Sigma_0 / \Sigma_\mathrm{b}$ is given by 
\begin{equation}
\frac{\Sigma_0}{\Sigma_\mathrm{b}} =1+\mathcal{A} \exp \left[-\frac{1}{2}\left(\frac{r-r_\mathrm{n}}{\Delta w} \right)^2 \right].
\end{equation}
The variation of $\Sigma_0 /\Sigma_\mathrm{b}$ is a gaussian bump whose height is $\mathcal{A}$ and width is $\Delta w$, and the peak of the bump is at $r=r_\mathrm{n}$. 
We call this the gaussian bump (GB) type. 
The GB type is intrinsically same as the ''HGB'' type in \citet{2000ApJ...533.1023L}. 
However, equation (7) is slightly different from the equation (4) in \citet{2000ApJ...533.1023L} about the definition of $\mathcal{A}$. 
\begin{figure}
\begin{center}
\epsscale{1.0}
\plotone{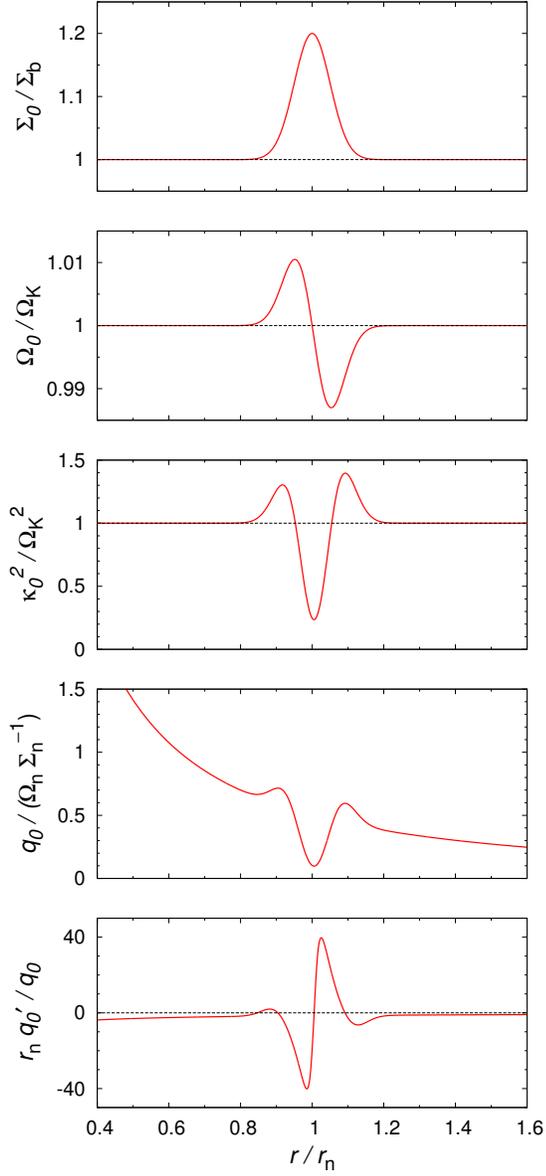}
\caption{Similar to Figure \ref{fig1}, but for the gaussian bump (GB) type with $\mathcal{A}=0.2$, $\Delta w =0.05 r_\mathrm{n}$, $h=0.1$ and $\Gamma=5/3$. 
The variation of $\Sigma_0 /\Sigma_\mathrm{b}$ is a gaussian bump whose height is $\mathcal{A}$ and width is $\Delta w$, and the peak of the bump is at $r=r_\mathrm{n}$. \label{fig3}}
\end{center}
\end{figure}

\item Gaussian Gap (GG)\\
$\Sigma_0 / \Sigma_\mathrm{b}$ is given by 
\begin{equation}
\frac{\Sigma_0}{\Sigma_\mathrm{b}} =(\mathcal{A}+1)-\mathcal{A} \exp \left[-\frac{1}{2}\left(\frac{r-r_\mathrm{n}}{\Delta w} \right)^2 \right].
\end{equation}
The variation of $\Sigma_0 /\Sigma_\mathrm{b}$ is a gaussian gap whose depth is $\mathcal{A}$ and width is $\Delta w$, and the minimum of the gap is at $r=r_\mathrm{n}$. 
We call this the gaussian gap (GG) type. 
\begin{figure}
\begin{center}
\epsscale{1.0}
\plotone{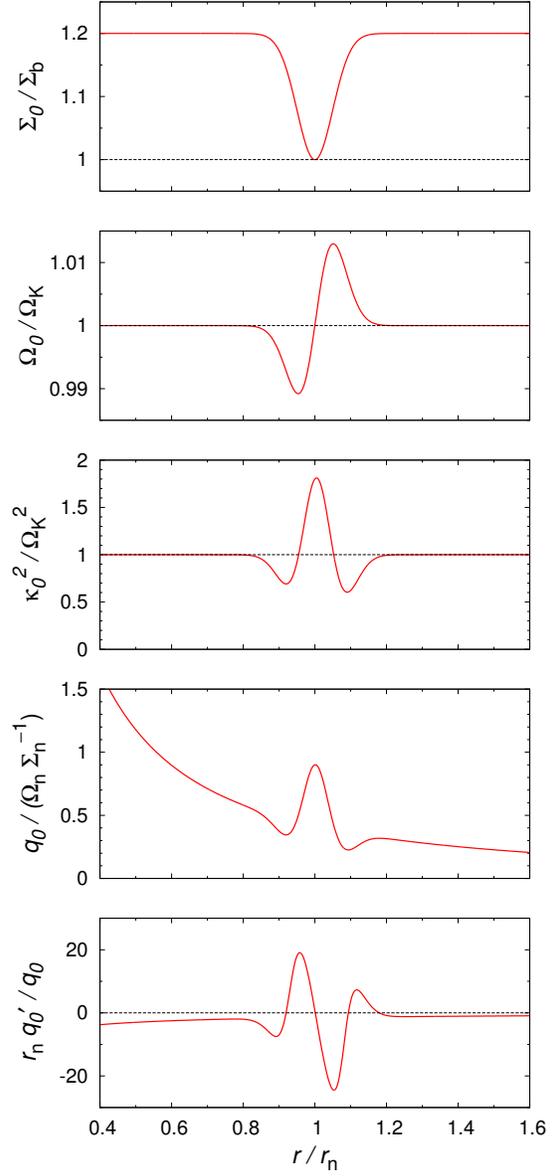}
\caption{Similar to Figure \ref{fig1}, but for the gaussian gap (GG) type with $\mathcal{A}=0.2$, $\Delta w =0.05 r_\mathrm{n}$, $h=0.1$ and $\Gamma=5/3$. 
The variation of $\Sigma_0 /\Sigma_\mathrm{b}$ is a gaussian gap whose depth is $\mathcal{A}$ and width is $\Delta w$, and the minimum of the gap is at $r=r_\mathrm{n}$.  \label{fig4}}
\end{center}
\end{figure}
\end{itemize}

Since we consider barotropic flow, the relation between $P_0 $ and $\Sigma_0 $ is written as
\begin{equation}
\frac{P_0}{P_\mathrm{n}} = \left(\frac{\Sigma_0}{\Sigma_\mathrm{n}}\right)^\Gamma ,
\end{equation}
where $P_\mathrm{n}$ is the (vertically integrated) pressure of the base disk model at $r=r_\mathrm{n}$. 
Since we consider the disk to be in vertically hydrostatic equilibrium in terms of the structure in the vertical ($z$) direction, the scale height $H_0 (r)$ is given by $H_0(r)= c_0/\Omega_0$, where $c_0 (r)\equiv \sqrt{\Gamma P_0/\Sigma_0}$ is background adiabatic sound speed and $\Omega_0 (r)\equiv v_{\varphi 0}/r$ is background rotation angular velocity. 
We define the dimensionless parameter $h$ as 
\begin{equation}
h\equiv \frac{\sqrt{\Gamma P_\mathrm{n}/\Sigma_\mathrm{n}}}{r_\mathrm{n} \Omega_\mathrm{n}}, 
\end{equation}
where $\Omega_\mathrm{K}(r) \equiv \sqrt{GM/r^3}$ is the Kepler angular rotation velocity and $\Omega_\mathrm{n} \equiv \Omega_\mathrm{K}(r_\mathrm{n})$ is the rotation velocity of the base disk model with $\mathfrak{f}=0$ at $r=r_\mathrm{n}$. 
The parameter $h$ equals the disk aspect ratio (scale height divided by radius) of the base model at $r=r_\mathrm{n}$. 
It is noted that $h$ can be regarded as the dimensionless scale height, or, equivalently, the dimensionless sound speed.
For background disks, equation (2) is rewritten as the radial forces balance;
\begin{equation}
\left( \frac{\Omega_0}{\Omega_\mathrm{K}} \right)^2 = 1 +h^2 \left(\frac{r}{r_\mathrm{n}} \right)^2 \left(\frac{\Sigma_0}{\Sigma_\mathrm{n}} \right)^{\Gamma-2} \left(\frac{\Sigma_0}{\Sigma_\mathrm{n}} \right)^\prime, 
\end{equation}
where the prime denotes $\mathrm{d} /\mathrm{d} r$. 

When $\Sigma_0$, $h$ and $\Gamma$ are given, it is possible to obtain $P_0$, $c_0$ and $\Omega_0$. 
The square of background epicyclic frequency $\kappa^2_0 (r)$; 
\begin{equation}
\kappa^2_0 (r) \equiv \frac{(r^4\Omega^2_0)^\prime}{r^3} 
\end{equation}
and background vortensity $q_0 (r)$;
\begin{equation}
q_0 (r)= \frac{(\mathrm{rot}\, {\bf v_0})_z}{\Sigma_0} = \frac{\kappa^2_0}{2\Sigma_0 \Omega_0} 
\end{equation}
can be also obtained. 

The background profile is characterized by five parameters, $(\mathcal{A}, \Delta w, h, \Gamma ,\mathfrak{f})$. 
In this paper, we fix $\mathfrak{f}=0$ and explore the parameter space for the four other quantities. 
Table 1 shows the parameter space explored in this study. 
We fix $h=0.1$ and $\Gamma=5/3$ and vary $\mathcal{A}$ and $\Delta w$ for the four types (SJU, SJD, GB, and GG) of background flow in cases (i)-(iii) and (vi). 
To see the effect of $h$ and $\Gamma$, we calculate $h=0.2$ case and $\Gamma=1$ case for the GB type in cases (iv) and (v) respectively. 
We note that $\Gamma=1$ indicates the case of isothermal disks. 
Figure \ref{fig1} shows $\Sigma_0 /\Sigma_\mathrm{b}$, $\Omega_0 /\Omega_\mathrm{K}$, $\kappa_0^2 /\Omega_\mathrm{K}^2$, $q_0$ and $q_0^\prime /q_0$ for the SJU type with $\mathcal{A}=0.5$, $\Delta w =0.05 r_\mathrm{n}$, $h=0.1$ and $\Gamma=5/3$. 
In like manner, Figure \ref{fig2} is for the SJD type with $\mathcal{A}=0.5$, Figure \ref{fig3} is for the GB type with $\mathcal{A}=0.2$, and Figure \ref{fig4} is for the GG type with $\mathcal{A}=0.5$. 
\begin{table}[t]
\begin{center}
\hspace{-3cm}
\caption{The parameters for each case.\label{tbl1}}
\begin{tabular}{ccccccccc}
\tableline\tableline
Case & Type & $h$ & $\Gamma$ & $\mathcal{A}$ & $\Delta w /r_\mathrm{n}$&$m$ \\
\tableline
i & SJU & 0.1 & 5/3 &0.001-10.0 &0.02-0.25 &1-10\\
ii & SJD & 0.1 & 5/3 &0.001-10.0 &0.02-0.25&1-10\\
iii & GB & 0.1 & 5/3 &0.001-10.0 &0.02-0.25&1-10\\
iv & GB & 0.1 & 1.0 &0.001-10.0 &0.02-0.25&1-10\\
v & GB & 0.2 & 5/3 &0.001-2.0 &0.02-0.25&1-10\\
vi & GG & 0.1 & 5/3 &0.001-10.0 &0.02-0.25&1-10\\
\tableline
\end{tabular}
\end{center}
\end{table}

\subsection{Perturbation Equations}
We perform linear stability analyses and use subscript "1" to denote perturbations, e.g., $\Sigma_1(r, \varphi, t)$, $P_1(r, \varphi, t)$ and ${\bf v}_1(r, \varphi, t)$, where ${\bf v}_1(r, \varphi, t)=v_{r1}(r, \varphi, t) {\bf \hat{r}} +v_{\varphi 1}(r, \varphi, t) {\bf \hat{\varphi}}$. 
The first-order forms of equations (1)-(3) are 
\begin{eqnarray}
\left( \henb{}{t}+\Omega_0 \henb{}{\varphi}\right) \Sigma_1+ \frac{1}{r}\henb{}{r} (r \Sigma_0 v_{r1})+\frac{\Sigma_0}{r}\henb{v_{\varphi 1}}{\varphi}=0, \nonumber\\
\left( \henb{}{t}+\Omega_0 \henb{}{\varphi}\right) v_{r1}-2\Omega_0 v_{\varphi 1}+\frac{1}{\Sigma_0}\henb{P_1}{r} - \frac{1}{\Sigma_0^2}\henb{P_0}{r} \Sigma_1= 0, \\
\left( \henb{}{t}+\Omega_0 \henb{}{\varphi}\right) v_{\varphi 1}+ \frac{\kappa^2_0}{2\Omega_0}v_{r1} +\frac{1}{r \Sigma_0}\henb{P_1}{\varphi}=0. \nonumber
\end{eqnarray}
Since we consider barotropic flow, 
\begin{equation}
P_1=c_{0}^2 \Sigma_1
\end{equation}
is satisfied. 
We consider perturbations as
\begin{eqnarray}
\Sigma_1(r, \varphi , t)=\Sigma_1(r) \exp (im\varphi -i\omega t), \nonumber\\
P_1(r, \varphi , t)=P_1(r) \exp (im\varphi -i\omega t), \nonumber\\
v_{r1}(r, \varphi , t)=v_{r1}(r) \exp (im\varphi -i\omega t), \nonumber\\
v_{\varphi 1}(r, \varphi , t)=v_{\varphi 1}(r) \exp (im\varphi -i\omega t), 
\end{eqnarray}
where $m(=1, 2, 3, $...) is the azimuthal mode number and $\omega$ is the mode frequency. 
In general, $\omega$ is a complex number and written as $\omega \equiv \omega_r+i\gamma$. 
The linearized forms of equations (14) are
\begin{eqnarray}
i \Delta \omega \Sigma_1=\left(\frac{\Sigma_0}{r}+ \Sigma_0^\prime \right)v_{r 1} +\Sigma_0 v_{r 1}^\prime +ik_\varphi \Sigma_0 v_{\varphi 1}, \\
i\Delta \omega v_{r1} +2\Omega_0 v_{\varphi 1} = \frac{P_1^\prime}{\Sigma_0}-\frac{P_0^\prime}{\Sigma_0^2}\Sigma_1, \\
i\Delta \omega v_{\varphi 1}-\frac{\kappa_0^2}{2\Omega_0}v_{r 1}=ik_\varphi \frac{P_1}{\Sigma_0}, 
\end{eqnarray}
where $\Delta \omega (r, \omega, m) \equiv \omega -m\Omega_0$ and $k_\varphi (r, m) \equiv m/r$. 

We define enthalpy perturbation as $\Psi (r) \equiv P_1/\Sigma_0$. 
It is noted that we omit the subscript ''1'' from $\Psi$ even though $\Psi$ is perturbation. 
Using $\Psi^\prime = P^\prime_1/\Sigma_0-\Psi \Sigma_0^\prime/\Sigma_0$, we obtain 
\begin{eqnarray}
i\Delta \omega v_{r1}+2\Omega_0 v_{\varphi 1} = \Psi^\prime, \\
i\Delta \omega v_{\varphi 1}-\frac{\kappa_0^2}{2\Omega_0}v_{r 1}=ik_\varphi \Psi
\end{eqnarray}
from equations (17)-(19). 
Solving equations (20)-(21) for $v_{r1}$ and $v_{\varphi 1}$, we have 
\begin{eqnarray}
\Sigma_0 v_{r1}=i \mathcal{F} \left(\frac{\Delta \omega}{\Omega_0}\Psi^\prime -2k_\varphi \Psi \right), \\
\Sigma_0 v_{\varphi 1}=\mathcal{F} \left( \frac{\kappa^2_0}{2\Omega^2_0}\Psi^\prime -k_\varphi \frac{\Delta \omega}{\Omega_0} \Psi \right) ,
\end{eqnarray}
where 
\begin{equation}
\mathcal{F}(r, \omega, m) \equiv \frac{\Sigma_0 \Omega_0}{\kappa^2_0 -\Delta \omega^2}.
\end{equation}
The perturbation equation for $\Psi$ is derived from equations (15), (17), (22)-(23) as 
\begin{equation}
\frac{1}{r}\left(\frac{r\mathcal{F}}{\Omega_0} \Psi^\prime \right)^\prime -\frac{k_\varphi^2 \mathcal{F}}{\Omega_0}\Psi =\left( \frac{\Sigma_0}{c_{0}^2} +\frac{2k_\varphi \mathcal{F}^\prime}{\Delta \omega} \right)\Psi . 
\end{equation}
This perturbation equation corresponds to the equation (14) in \citet{2000ApJ...533.1023L}. 
Equation (25) can be written as 
\begin{equation}
\Psi^{\prime \prime} + B(r, \omega, m)\Psi^\prime +C(r, \omega, m)\Psi =0, 
\end{equation}
where 
\begin{eqnarray}
B(r, \omega, m) \equiv \frac{1}{r}+\frac{\mathcal{F}^\prime}{\mathcal{F}}-\frac{\Omega_0^\prime}{\Omega_0}, \\
C(r, \omega, m) \equiv -k_\varphi^2-\frac{\kappa^2_0-\Delta \omega^2}{c_{0}^2}-2k_\varphi \frac{\Omega_0}{\Delta \omega}\frac{\mathcal{F}^\prime}{\mathcal{F}}.
\end{eqnarray}
When we define 
\begin{equation}
\Xi (r) \equiv \sqrt{r\frac{\mathcal{F}}{\Omega_0}}\Psi ,
\end{equation}
equation (26) can be rewritten as
\begin{equation}
\Xi^{\prime \prime} -D(r, \omega, m)\Xi =0, 
\end{equation}
where 
\begin{equation}
D(r, \omega, m)\equiv \frac{B^\prime}{2}+\frac{B^2}{4}-C . 
\end{equation}
Equation (30) is similar to the stationary Schr\"odinger equation with zero energy, and $D(r, \omega, m)$ can be regarded as the effective potential. 
The region where $\mathrm{Re}[D]>0$ is an evanescent region and the region where $\mathrm{Re}[D]<0$ is a wave propagation region \citep{1979ApJ...233..857G}. 
We also omit the subscript ''1'' from $\Xi$ even though $\Xi$ is perturbation. 

We solve equation (26) or (30) under appropriate boundary conditions (see \S 2.4) at the outer and inner radii of the disk. 
We obtain the eigenvalue $\omega = \omega_r + i \gamma$, where $\omega_r$ and $\gamma$ are the real and the imaginary part of the eigenvalue, respectively.  
From the real part of the eigenvalue, we can calculate the co-rotation radius $r_\mathrm{c}$ by solving $\Omega_0(r_\mathrm{c}) = \omega_r/m$. 
The imaginary part of the eigenvalue gives the growth rate of the mode against to the RWI if $\gamma$ is positive. 
For $\gamma=0$, the mode is marginally stable to the RWI. 
We obtain $\Psi(r)$ or $\Xi(r)$ as well as the mode frequency $\omega$. 
From $\Psi(r)$ or $\Xi(r)$, other perturbations (e.g., $\Sigma_1, P_1, v_{r1}, v_{\varphi 1}$) are evaluated. 
To solve equation (26) or (30), we discretized the equation in the radial direction with $N$ grids and use matrix inversion (see Appendix A for details). 
The inner and outer boundaries are taken at $r_\mathrm{inner}=0.3 r_\mathrm{n}$ and $r_\mathrm{outer} = 3.0 r_\mathrm{n}$. 
We adopt $N=3000$ unless otherwise stated. 

\begin{figure}
\begin{center}
\epsscale{1.0}
\plotone{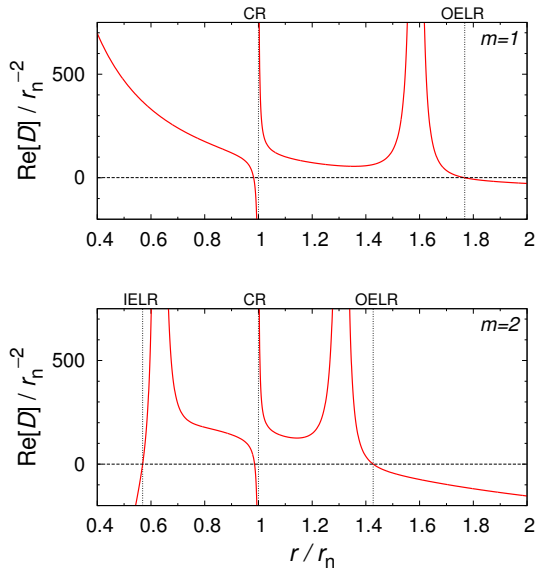}
\caption{
The real part of the effective potential $\mathrm{Re}[D]$ for the base disk model with $h=0.1$, $\Gamma=5/3$, $\mathfrak{f}=0$ and $\omega/ \Omega_\mathrm{n}=(m, 10^{-5})$. 
Top panel shows the $m=1$ case and bottom panel shows the $m=2$ case. 
Since the base disk model has the Keplerian rotation profile, the co-rotation radius is located at $r=r_\mathrm{n}$. 
CR denotes the co-rotation radius. 
$\mathrm{Re}[D]$ has both the IELR and the OELR for $m=2$, while only the OELR for $m=1$. 
\label{fig5}}
\end{center}
\end{figure}

Figure \ref{fig5} shows the real part of the effective potential $\mathrm{Re}[D]$ for the base disk model with $\mathfrak{f}=0$ and $r_\mathrm{c}=r_\mathrm{n}$ in the cases of $m=1$ (top panel) and $m=2$ (bottom panel). 
Turning points of $\mathrm{Re}[D]$, where $\mathrm{Re}[D]=0$ is satisfied, except in the vicinity of the co-rotation radius are the effective Lindblad resonances \citep[ELRs;][]{1993ApJ...419..155A}. 
Outer one is the outer effective Lindblad resonance (OELR), and inner one is the inner effective Lindblad resonance (IELR). 
It is noted that the IELR does not exist for $m=1$. 
The case for higher azimuthal mode number ($m=3, 4...$) is similar to the case for $m=2$, but the ELRs are closer to the co-rotation radius. 
For the base disk models where the radial surface mass density distribution is given by a simple power-law profile, the region between the ELRs is an evanescent region except in the vicinity of the co-rotation radius. 
When background profiles have a rapid radial variation, the region between the ELRs is not always an evanescent region (see \S 5.1). 
On the other hand, the regions inside the IELR and outside the OELR are always wave propagation regions. 

\subsection{Boundary Conditions}

To solve equation (26) or (30), appropriate boundary conditions should be set. 
We adopt the outgoing boundary conditions. 
The region that is sufficiently far away from the transition of $\Sigma_0$, $r=r_\mathrm{n}$, is a wave propagation region except the inner region for $m=1$. 
We first consider boundary conditions except in the case of the inner boundary for $m=1$. 
When we choose the inner boundary inside the IELR and the outer boundary outside the OELR, we can use the WKBJ solution of the form $\Psi (r)\propto \exp [i \mathfrak{g}(r)]$. 
When we substitute this into equation (26) and neglect the term which includes $\mathfrak{g}^{\prime \prime}$, 
\begin{equation}
\mathfrak{g}^{\prime~ 2}-iB\mathfrak{g}^\prime -C=0 
\end{equation}
is obtained. 
From equation (32), $\mathfrak{g}^\prime$ is solved as 
\begin{equation}
\mathfrak{g}^\prime (r) =\frac{iB\pm \sqrt{4C-B^2}}{2}. 
\end{equation}
The positive sign in equation (33) indicates the trailing, or outgoing, wave. 
Therefore, we apply 
\begin{equation}
\Psi^\prime=\frac{-B(r)+ i \sqrt{4C-B^2}}{2} \Psi 
\end{equation}
as the boundary conditions on $\Psi (r)$ except in the case of the inner boundary for $m=1$. 

For $m=1$, the inner boundary falls in the evanescent region. 
The perturbation $\Psi$ is required to be regular at $r=0$. 
We assume that the solution is of the form of $\Psi (r)\propto r^\mathfrak{h}$. 
From equations (27)-(28), we have $B\propto 1/r$ and $C\propto 1/r^2$ for $r\sim 0$. 
From equation (26), $\mathfrak{h}$ is solved as 
\begin{equation}
\mathfrak{h}=\frac{(1-rB)\pm \sqrt{(rB-1)^2-4r^2C}}{2}. 
\end{equation}
Since $\Psi$ is regular at $r=0$, we have to choose positive sign in equation (35). 
Therefore, we apply 
\begin{equation}
\Psi^\prime =\frac{1}{2}\left\{-\left[ B-\frac{1}{r}\right]+\sqrt{\left[ B-\frac{1}{r}\right]^2-4C}\right\} \Psi 
\end{equation}
as the boundary condition on $\Psi(r)$ in the case of the inner boundary for $m=1$.

In the same way, boundary conditions on $\Xi (r)$ are also obtained from equation (30) as 
\begin{equation}
\Xi^\prime = i \sqrt{-D} \Xi
\end{equation}
except in the case of the inner boundary for $m=1$ and 
\begin{equation}
\Xi^\prime =\left[\frac{1}{2r}+\sqrt{\frac{1}{4r^2}+D} \right]\Xi
\end{equation}
in the case of the inner boundary for $m=1$.

\section{Associated Known Conditions with the RWI}
There are two known conditions associated with the RWI. 
Before presenting the results of calculations, we briefly review these known conditions. 
One is the Rayleigh's condition, which is the condition for the rotational instability to occur \citep[e.g.,][]{1960PNAS...46..253C}. 
The other is the Lovelace's condition \citep{1999ApJ...513..805L}. 

\subsection{Rayleigh's Condition}
When there is the radius where 
\begin{equation}
\kappa^2_0(r) <0 
\end{equation}
is satisfied, the system is unstable to the rotational instability, which is a axisymmetric hydrodynamical instability in differentially rotating disks \citep{1960PNAS...46..253C}. 
Equation (39) is known as the Rayleigh's criterion. 
We call the condition that there is the radius where the Rayleigh's criterion is violated the ''Rayleigh's condition''. 
\citet{2000ApJ...533.1023L} indicates that the Rayleigh's condition is expected to be the sufficient condition for the RWI. 

\subsection{Lovelace's Condition}
We multiply equation (25) by $r\Psi^\ast$ and integrate over the disk, where $\Psi^\ast$ is the complex conjugate of $\Psi$. 
Assuming $r\mathcal{F}|\Psi|^2/\Omega_0 \rightarrow 0$ for $r\rightarrow 0, \infty$, we obtain 
\begin{equation}
\int \mathrm{d}r\, \left[\frac{r\mathcal{F}}{\Omega_0}(|\Psi^\prime|^2+k_\varphi^2|\Psi|^2) +r\left( \frac{\Sigma_0}{c_0^2}+\frac{2k_\varphi \mathcal{F}^\prime}{\Delta \omega}\right)|\Psi|^2 \right]=0. 
\end{equation}
Assuming that the dominant contribution to the integral of the left hand side of equation (40) comes from the region in the vicinity of the co-rotation radius, it is expected that $\Delta \omega^2 \ll \kappa_0^2$ and $\mathcal{F} \sim 1/2q_0$. 
Then, the imaginary part of equation (40) can be written as 
\begin{equation}
\int \mathrm{d}r\, \left[\frac{m |\Psi|^2}{q_0^2} \, \mathrm{Im}\left(\frac{1}{\Delta \omega}\right) q_0^\prime \right] =0. 
\end{equation}
When the system is unstable to the RWI and stable to the rotational instability, $[m |\Psi|^2 \, \mathrm{Im}\left(1/\Delta \omega \right)/q_0^2]$ is always negative. 
To satisfy equation (41), the existence of a turning point in the profile of $q_0^\prime$ is required, i.e., background vortensity $q_0$ should have at least one extremum if the disk is unstable to the RWI \citep{1999ApJ...513..805L}. 
The Lovelace's condition is considered as the necessary condition for the RWI. 

\section{Most Unstable States to the RWI} 
For each set of background flow parameters $(\mathcal{A}, \Delta w, h, \Gamma)$, we calculate the most unstable mode frequency $\omega_\ast=\omega_{r\ast}+i\gamma_\ast$ and the most unstable azimuthal mode number $m_\ast$ to the RWI. 
The largest growth rate $\gamma_\ast$ is the growth rate of the system against the RWI if $\gamma_\ast$ is positive. For $\gamma_\ast =0$, the system is marginally stable to the RWI. 

\begin{figure*}
\epsscale{2.0}
\plotone{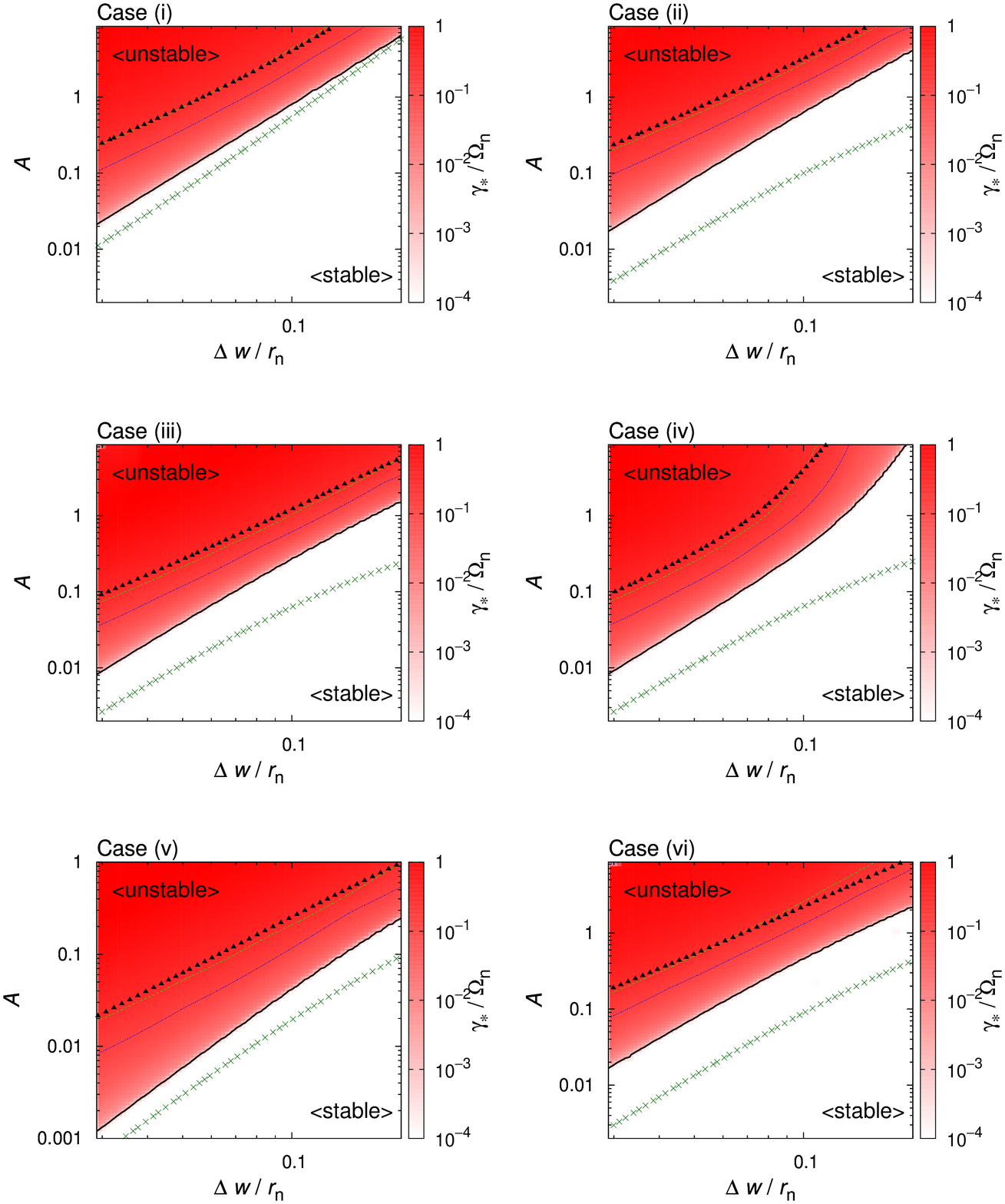}
\vspace{-2cm}
\caption{
The largest growth rate $\gamma_\ast$ is shown as color in the $\Delta w$ - $\mathcal{A}$ plane for each case as seen Table 1. 
The bottom-right white region corresponds to the parameters for the stable states of the system. 
The black solid line is the contour for $\gamma_\ast /\Omega_\mathrm{n} = 10^{-3}$, the blue dotted line for $\gamma_\ast /\Omega_\mathrm{n} = 0.1$ and the light-green dashed line for $\gamma_\ast /\Omega_\mathrm{n} = 0.2$. 
The line of dark-green cross represents the marginal states of the Lovelace's condition, and the region above this line is where the Lovelace's condition is satisfied. 
The line of  black triangle represents the marginally stable states to the rotational instability, and the region above this line is where the Rayleigh's condition is violated. \label{fig6}}
\end{figure*}

We first show the largest growth rate $\gamma_\ast$ with every $\mathcal{A}$ and $\Delta w$ for representative 6 cases in Figure \ref{fig6}. 
In each panel of Figure \ref{fig6}, we also show the parameters for marginally stable states of the system to the RWI. 
Considering numerical errors (see Appendix A), we determine that the system is ''nearly'' marginally stable to the RWI if $\gamma_{\ast}/\Omega_\mathrm{n}$ is no more than $10^{-3}$. 
We note that there is no significant difference in the parameters for the marginally stable states of the system if we decrease the threshold to $10^{-4}$. 
The methods of calculations to derive parameters exclusively for the marginally stable states will be presented in \S 5.2, but we obtain essentially the same results for the parameters of the marginally stable states (see \S 5.3 for details). 

For all the cases (i)-(vi), the largest growth rates are similar, but we see that the system can be more unstable for case (v), where $h$ is twice as large as other cases. 
However, we have found that all the cases look very similar to each other if we use $\Delta w /(h r_\mathrm{n})$ in the horizontal axis in Figure \ref{fig6} instead of $\Delta w/r_\mathrm{n}$. 
Therefore, it is expected that doubling $h$ with fixed $\Delta w$ has the same effect as halving $\Delta w$ with $h$ fixed. 

In Figure \ref{fig6}, we show the parameter space where the Rayleigh condition is not satisfied but Lovelace's condition is violated. 
It is clear that the marginally stable parameters of the system against the RWI fall in this region. 
Therefore, we can regard the Lovelace's condition as the necessary condition for the RWI and the Rayleigh's condition as the sufficient condition for the RWI at least within the parameter space which we have explored. 
The sufficiency of the Rayleigh's condition, however, has more issues, and we actually condsider that this does not always hold.  Further discussions will be presented in Section 5.6. 
In addition, we can know that the Lovelace's condition is not the sufficient condition for the RWI. 
\begin{figure*}
\epsscale{2.0}
\plotone{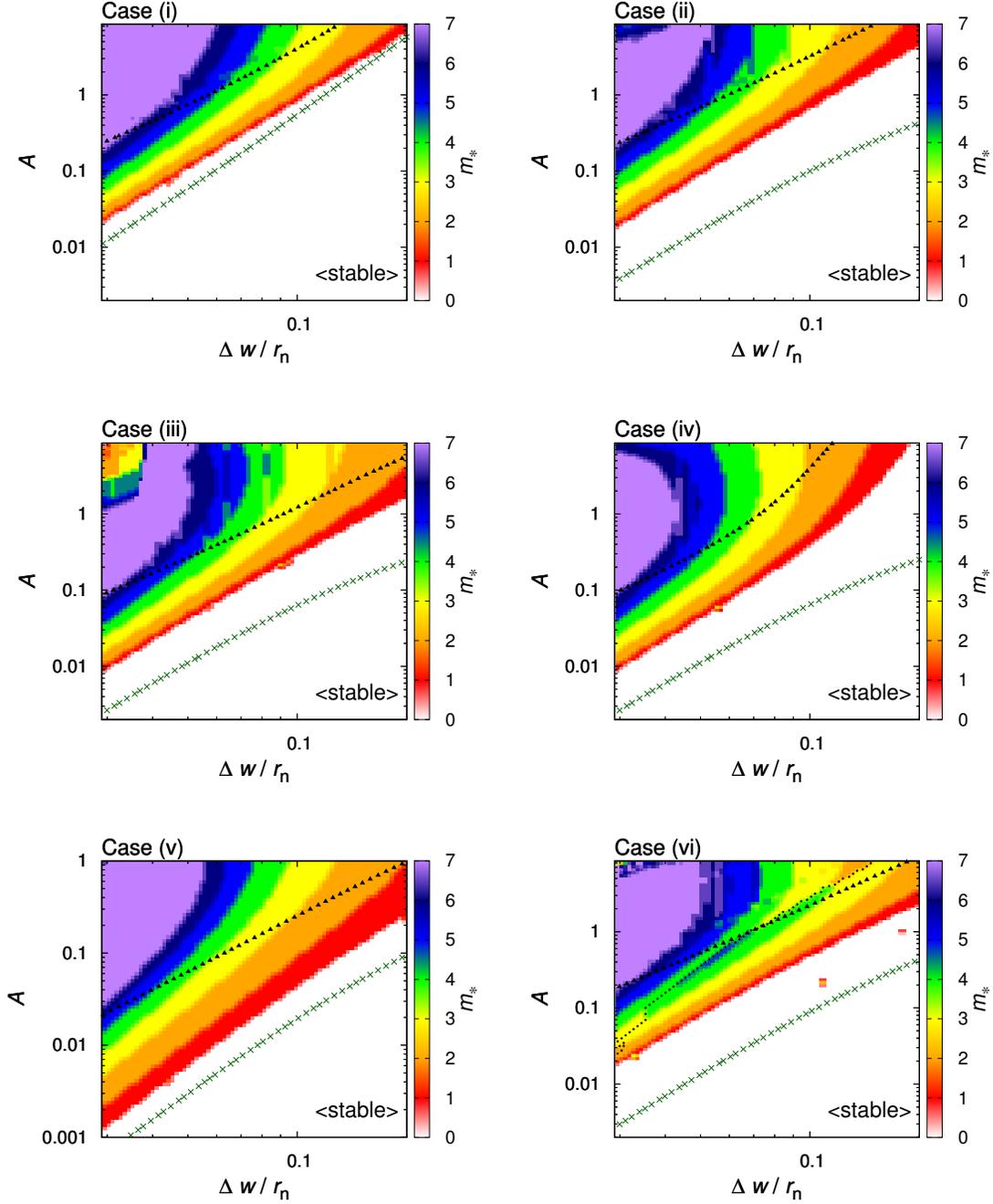}
\vspace{-2cm}
\caption{Similar to Figure \ref{fig6}, but the most unstable mode $m_\ast$ is shown as color (red: 1, orange: 2, yellow: 3, green: 4, blue: 5, dark-blue: 6, purple: no less than 7). 
The $m=0$ region corresponds to the parameters for the stable states of the system. 
The black dotted line for the case (vi) is the transition line of the co-rotation radius. \label{fig7}}
\end{figure*}
\begin{figure*}
\epsscale{2.0}
\plotone{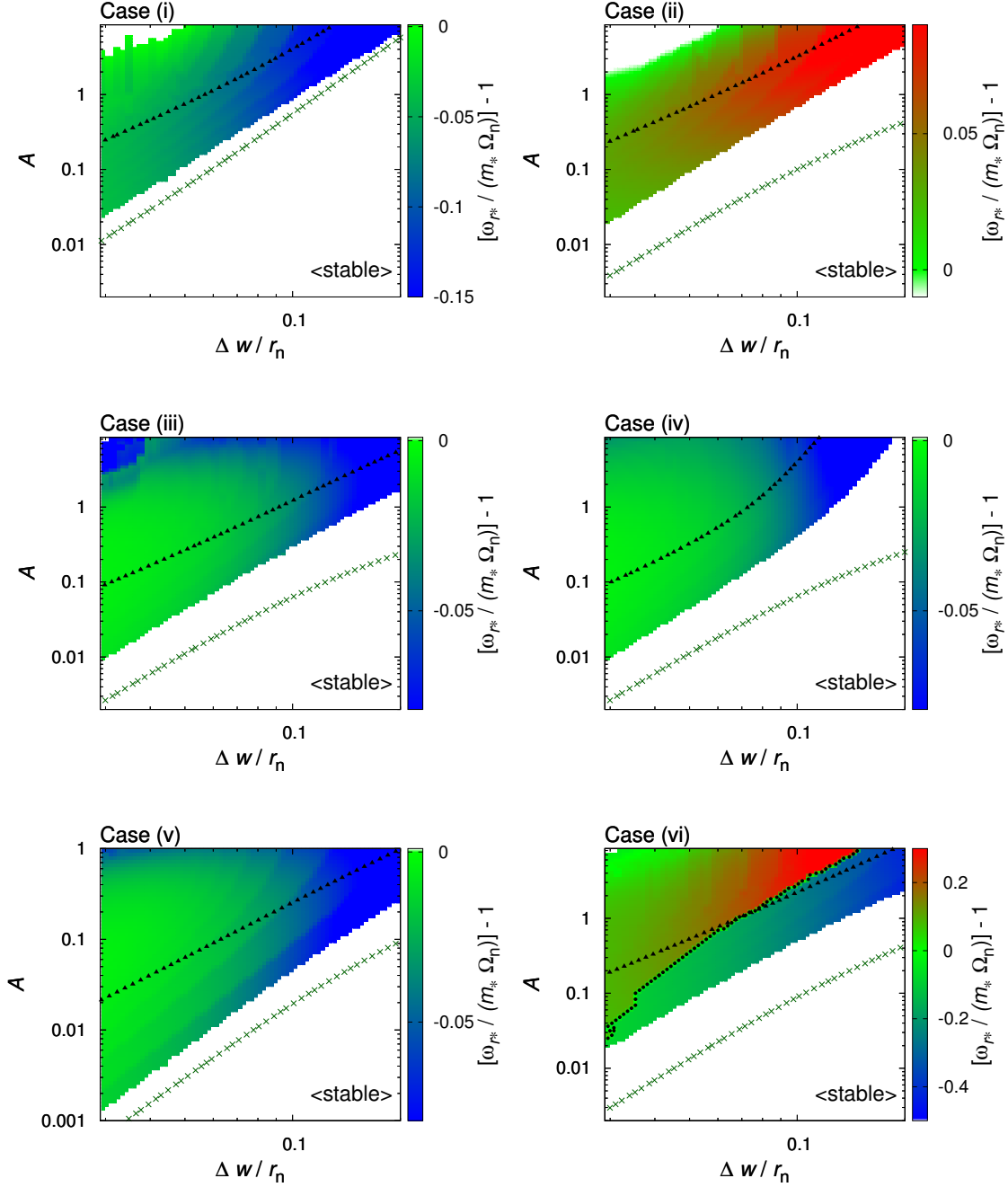}
\vspace{-2cm}
\caption{Similar to Figure \ref{fig6}, but $\omega_{r\ast}/m_\ast$ is shown as color. 
Due to $\Omega_0^\prime <0$, the co-rotation radius is located the inside of $r=r_\mathrm{n}$ for $\omega_{r\ast}/(m_\ast \Omega_\mathrm{n})>1$, and vice versa. 
The black dotted line for the case (vi) is the transition line of the co-rotation radius.
 \label{fig8}}
\end{figure*}

In Figure \ref{fig7}, we show the most unstable azimuthal mode number $m_{\ast}$ for various background parameters. 
For the parameters where the rotational instability is stable, $m_\ast$ is no lager than $7$, and we describe $m_\ast \geq 7$ in one color (purple). 
We define that the "system is stable" if there is no mode number that gives positive growth rate $\gamma$.
From Figure \ref{fig7}, the less $m_\ast$ is, the closer to the marginally stable states the system is. 
If the system is close to the marginally stable states, $m_\ast$ is unity. 
The general behavior of $m_{\ast}$ is similar in all the six cases. 

In Figure \ref{fig8}, we show $\omega_{r\ast}/m_\ast$ for various background parameters. 
From the results of calculations, we can obtain the co-rotation radius $r_\mathrm{c}$  for each most unstable mode by solving $\Omega_0(r_\mathrm{c})= \omega_{r\ast}/m_\ast$. 
Due to $\Omega_0^\prime <0$, the co-rotation radius is located inside $r=r_\mathrm{n}$ if $\omega_{r\ast}/(m_\ast \Omega_\mathrm{n})$ is larger than unity, and vice versa. 
In general, the co-rotation radius is located close to (but not exactly at the same radius as) the background vortensity ($q_0$) minimum with large concavity ($q_0^{\prime \prime}$). 
If there are several $q_0$ minima with similar concavity, the co-rotation radius is located in the vicinity of one of the $q_0$ minima. 
For the GG type, there are two $q_0$ minima with large $q_0^{\prime \prime}$. 
The co-rotation radius actually changes from outside $r=r_\mathrm{n}$ to inside $r=r_\mathrm{n}$ at the black dotted line with the increase of $\mathcal{A}$ for case (vi) in Figure \ref{fig8}. 
For other types, there is only one $q_0$ minimum with large $q_0^{\prime \prime}$, and the region which has the co-rotation radius is uniquely determined. 
We note that $m_\ast$ is also changed at the transition line of the co-rotation radius for case (vi) in Figure \ref{fig7}. 

Interestingly, in Figure \ref{fig6}, the line with $\gamma_{\ast}/\Omega_\mathrm{n} = 0.2$ overlaps with the marginally stable line to the rotational instability. 
This may indicate the existence of some physical relationship between the rotational instability and the RWI. 
However, we focus on the marginally stable states to the RWI and do not discuss further about the RWI in the regime where the Rayleigh condition is violated. 
The effective potential $D$ is very complicated in the regime where the Rayleigh's condition is strongly violated, and it is difficult to find the largest growth rate accurately in such a parameter regime. 
This issue should be discussed elsewhere. 

\section{Marginally Stable States to the RWI and Conditions for the RWI}
\subsection{Nearly Marginally Stable States as Derived in Our Calculation}
In the previous section, we have shown the mode frequencies and the azimuthal mode numbers for the most unstable states to the RWI. 
We now investigate, in more detail, the marginally stable states. 
The key to understand the RWI is the effective potential $D$, which is introduced in equation (30). 
We focus on the calculations with case (iii) ($h=0.1$, $\Gamma=5/3$, the GB type) with $\Delta w = 0.05 r_\mathrm{n}$ and $m=1$ or $2$. 
For each azimuthal mode number ($m=1$ or $2$), the growth rate $\gamma$ can be determined if we specify the amplitude of the bump $\mathcal{A}$. 
To put it in the other way around, there is an amplitude of the bump $\mathcal{A}$ that gives the specified growth rate $\gamma$. 
Therefore, in this section, we specify several values of the growth rate $\gamma$ and compare the effective potentials and the eigenfunctions.

\begin{figure*}
\begin{center}
\epsscale{2.0}
\plotone{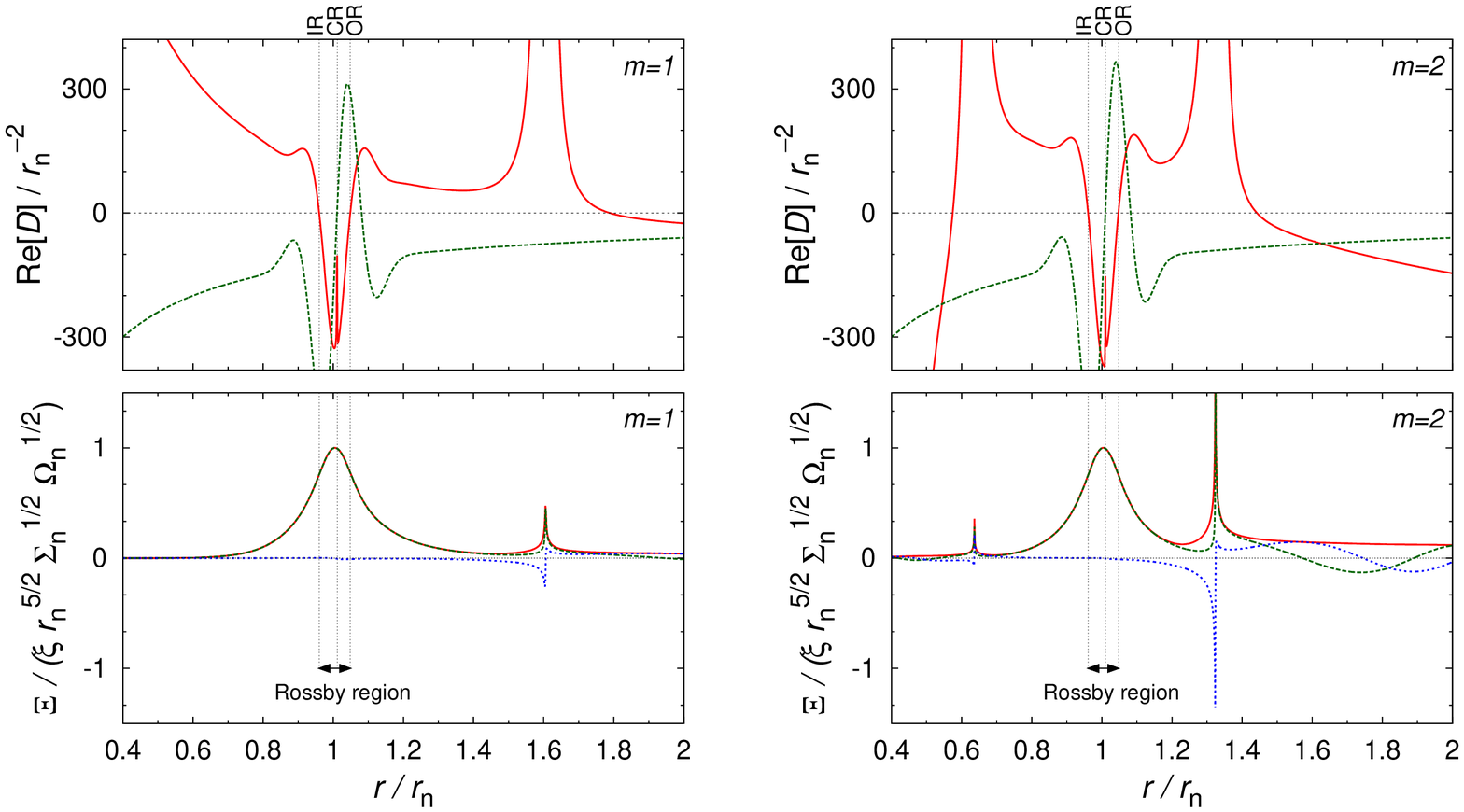}
\caption{
The real part of the effective potential $\mathrm{Re}[D]$ (top panels) and the eigenfunction $\Xi$ (bottom panels) are shown for $\gamma /\Omega_\mathrm{n}\simeq 10^{-3}$ and case (iii) with $\Delta w =0.05 r_\mathrm{n}$. 
This corresponds to the case for $\mathcal{A}=0.04$, $\omega_r /\Omega_\mathrm{n}= 0.983 \times 1$ and $r_\mathrm{c}=1.011 r_\mathrm{n}$ for $m=1$ (left panels), and $\mathcal{A}=0.04473$, $\omega_r /\Omega_\mathrm{n}=0.985 \times 2$ and $r_\mathrm{c}=1.010 r_\mathrm{n}$ $m=2$ (right panels). 
In top panels, $\mathrm{Re}[D]$ and $q_0^\prime /q_0$ are represented with the red solid line and the green dashed line. 
In bottom panels, $|\Xi |$, $\mathrm{Re}[\Xi ]$ and $\mathrm{Im}[\Xi ]$ are represented with the red solid line, the green dashed line and the blue dotted line. 
IR denotes the inner Rossby turning point, OR does the outer Rossby turning point and CR does the co-rotation radius. 
The region between IR and OR is the Rossby region. 
We note that $\xi$ is defined as the normalization constant of the perturbation by $\xi \equiv \mathrm{Re}[\Xi (r=r_\mathrm{p})]/(r_\mathrm{n}^{5/2} \Sigma_\mathrm{n}^{1/2} \Omega_\mathrm{n}^{1/2})$, where $r_\mathrm{p}$ is the radius where $\Xi$ exhibits the maximum within the Rossby region. 
We choose $\xi=0.025$ throughout this paper. \label{fig9}}
\end{center}
\end{figure*}

Figure \ref{fig9} shows the real part of the effective potential $\mathrm{Re}[D]$ (top panels) and $\Xi$ (bottom panels) for $\gamma /\Omega_\mathrm{n}\simeq 10^{-3}$. 
These cases correspond to $\mathcal{A}=0.04$ for $m=1$ and $\mathcal{A}=0.04473$ for $m=2$. 
The real part of the eigenvalue or the co-rotation radius is $\omega_r /\Omega_\mathrm{n}= 0.983 \times 1$ or $r_\mathrm{c}=1.011 r_\mathrm{n}$ for $m=1$ and $\omega_r /\Omega_\mathrm{n}=0.985\times 2$ or $r_\mathrm{c}=1.010 r_\mathrm{n}$ for $m=2$. 
From Figure \ref{fig9}, $\mathrm{Re}[D]$ has three ($m=1$) or four ($m=2$) turning points, where $\mathrm{Re}[D]=0$ is satisfied, except in the vicinity of the co-rotation radius. 
Compared to the case for the base disk model without a bump (Figure \ref{fig5}), there are two additional turning points other than ELRs. 
Here, we call the new turning points ''Rossby turning points'', and the region between Rossby turning points ''Rossby region''. 
Recalling that $D$ is considered as the effective potential of the stationary Schr\"odinger equation with zero energy, the Rossby region is a wave propagation region. 
The Rossby region is regarded as a potential well and therefore the real part of the eigenfunction $\mathrm{Re}[\Xi ]$ is trapped in the Rossby region. 
The co-rotation radius is very close to the  radius of the $q_0$ minimum. 
In producing the bottom panels of Figure 9, it is necessary to define a normalization constant of the perturbation with respect to the background. 
In this paper, we define the normalization constant of the perturbation by $\xi \equiv \mathrm{Re}[\Xi (r=r_\mathrm{p})]/(r_\mathrm{n}^{5/2} \Sigma_\mathrm{n}^{1/2} \Omega_\mathrm{n}^{1/2})$, where $r_\mathrm{p}$ is the radius where $\Xi$ exhibits the maximum within the Rossby region. 
We choose $\xi=0.025$ throughout this paper. 
\begin{figure}
\begin{center}
\epsscale{1.0}
\plotone{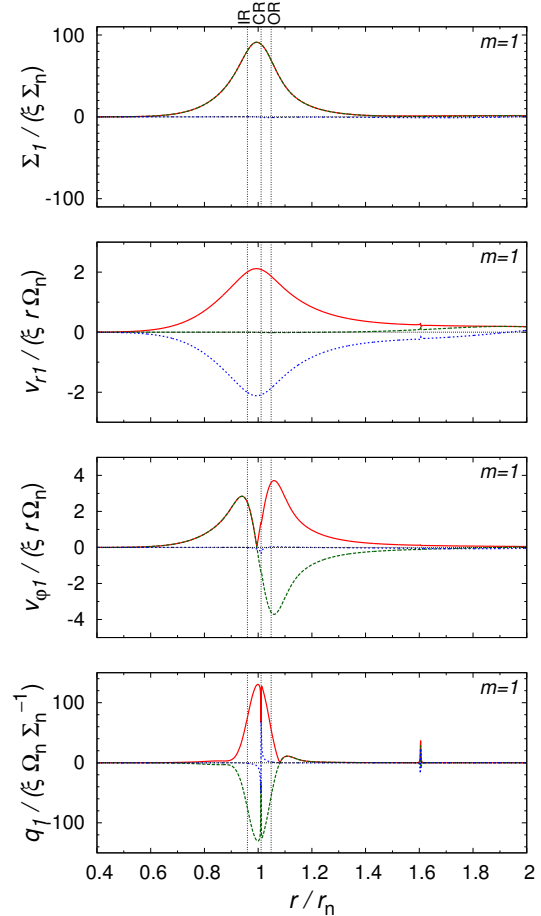}
\caption{Perturbations ($\Sigma_1$, $v_{r1}$, $v_{\varphi 1}$, $q_1$) are shown for the same case as the left panels in Figure \ref{fig9}. 
Absolute values, real parts and imaginary parts of perturbations are represented with the red solid line, the green dashed line and the blue dotted line. 
IR denotes the inner Rossby turning point, OR does the outer Rossby turning point and CR does the co-rotation radius. 
 \label{fig10}}
\end{center}
\end{figure}

Figure \ref{fig10} shows perturbations ($\Sigma_1$, $v_{r1}$, $v_{\varphi 1}$, $q_1$), where  
\begin{equation}
q_1 (r)\equiv \frac{1}{\Sigma_0}\left[ \frac{1}{r}(rv_{\varphi 1})^\prime-ik_\varphi v_{r1}-q_0 \Sigma_1\right] 
\end{equation}
is vortensity perturbation, for the same case as the left panels in Figure \ref{fig9} ($\gamma /\Omega_\mathrm{n}\simeq 10^{-3}$ for $m=1$). 
The imaginary parts of $\Sigma_1$, $v_{\varphi 1}$ and $q_1$ and the real part of $v_{r1}$ are almost equal to zero. 
The real part of $\Sigma_1$ and the imaginary part of $v_{r 1}$ have no node and peaks in the vicinity of the co-rotation radius. 
Meanwhile, the real part of $v_{\varphi 1}$ has a node in the vicinity of the co-rotation radius. 
The real part of $q_1$ also has no node like $\mathrm{Re} [\Sigma_1 ]$ and $\mathrm{Im}[v_{r1}]$, but a spike at the co-rotation radius. 
Unlike other perturbations, $\mathrm{Re}[q_1]$ is almost confined within the Rossby region. 
Perturbations have similar trends for $m=2$ to those for $m=1$. 

\begin{figure*}
\epsscale{2.0}
\plotone{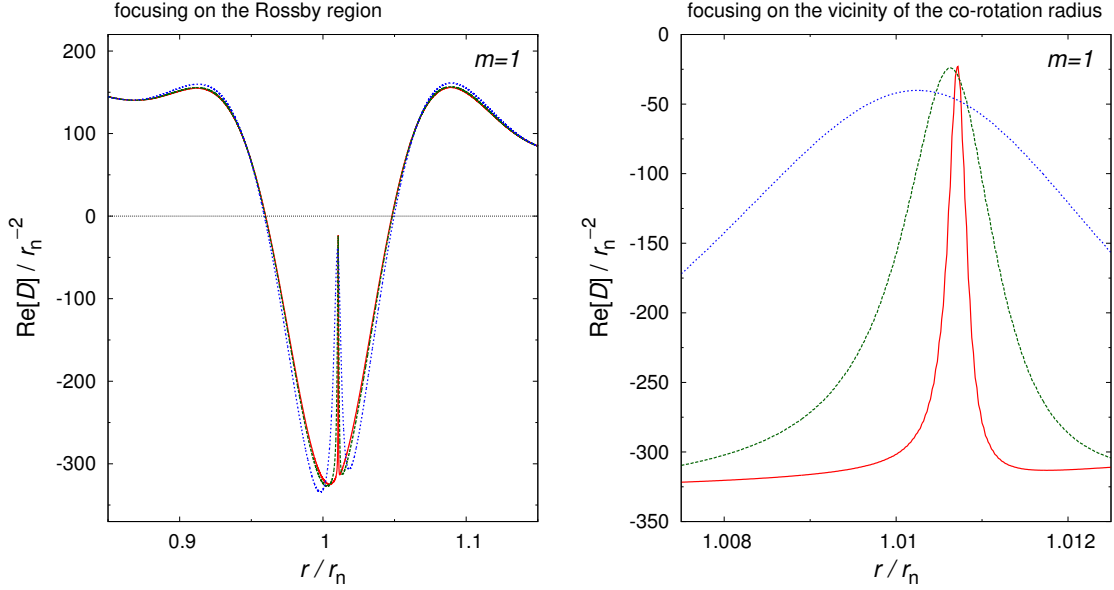}
\caption{
$\mathrm{Re}[D]$ is shown for $\gamma /\Omega_\mathrm{n} \simeq 2\times 10^{-4}$ ($\mathcal{A}=0.03920)$ with red solid line, $10^{-3}$ ($\mathcal{A}=0.04$) with green dashed line and $5\times 10^{-3}$ ($\mathcal{A}=0.04422$) with blue dotted line for case (iii) with $\Delta w =0.05 r_\mathrm{n}$. 
We focus on the Rossby region in left panel and the vicinity of the co-rotation radius in right panel. 
It is noted that we adopt the grid $N=20000$ in order to resolve the vicinity of the co-rotation radius. 
\label{fig11}}
\end{figure*}

To look into the change of $\mathrm{Re}[D]$ with varying $\mathcal{A}$ for $m=1$, Figure \ref{fig11} shows $\mathrm{Re}[D]$ for $\gamma /\Omega_\mathrm{n} \simeq 2\times 10^{-4}$ ($\mathcal{A}=0.03920$), $10^{-3}$ ($\mathcal{A}=0.04$) and $5\times 10^{-3}$ ($\mathcal{A}=0.04422$). 
We focus on the Rossby region in the left panel and on the vicinity the co-rotation radius in the right panel. 
It should be noted that $\mathrm{Re}[D]$ almost diverges at the co-rotation radius since $C$ defined in equation (28) has a term proportional to $1/ \Delta \omega$. 
However, as seen in Figure \ref{fig11}, the width of the singularity at the co-rotation radius becomes narrower while the peak does not change very much as we approach to the marginally stable states. 
This indicates that the singularity at the co-rotation radius becomes weak as we approach to the marginally stable states where $\gamma=0$. 
It should be noted that we adopt $N=20000$ in Figure \ref{fig11} in order to resolve $\mathrm{Re}[D]$ in the vicinity of the co-rotation radius. 
The change of $\mathrm{Re}[D]$ with varying $\mathcal{A}$ is also similar for $m=2$ to that for $m=1$. 
\begin{figure}
\begin{center}
\epsscale{0.9}
\plotone{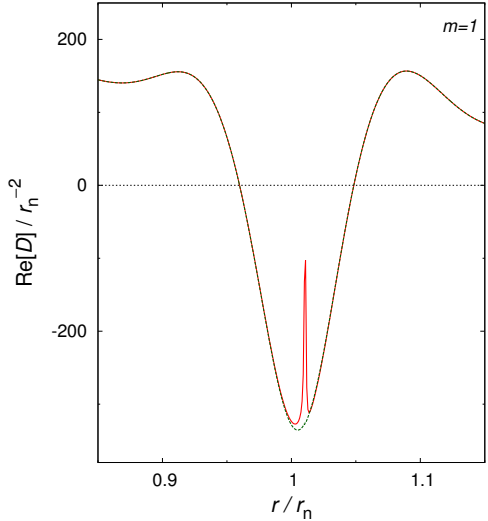}
\caption{$\mathrm{Re}[D]$ (red solid line) and $D_\mathrm{MS}$ (green dashed line) are shown for the same case as the left panels in Figure \ref{fig9}. \label{fig12}}
\end{center}
\end{figure}

\subsection{Mode Frequency and Effective Potential for the Marginally Stable States}
It is actually possible to derive the fact that there is no singularity at the co-rotation radius for the marginally stable states. 
For barotropic flow, vortensity is always conserved with respect to fluid elements as equation (4). 
The linearized form of equation (4) is written as 
\begin{equation}
i\Delta \omega \, q_1 =v_{r1}q_0^\prime .
\end{equation}
Since we have $\Omega_0 = \omega_r(r_\mathrm{c})/m$ and $\gamma=0$, $\Delta \omega$ is equal to zero at the co-rotation radius for the marginally stable states. 
In other words, either $v_{r1}$ or $q_0^\prime$ should be equal to zero at the co-rotation radius for the marginally stable states. 
From Figure \ref{fig10}, it is expected that $\mathrm{Im}[v_{r1}]$ has no node then. 
Therefore, the co-rotation radius has to locate at the radius of the $q_0$ extremum for the marginally stable states. 
The co-rotation radius is located at the radius of the $q_0$ minimum rather than one of the $q_0$ maximum in fact. 
This will be discussed in \S 6. 

For the marginally stable states, we have $\omega_r = m \Omega_\mathrm{0}(r_\mathrm{v})$, where $r_\mathrm{v}$ is the radius of the $q_0$ minimum; 
\begin{equation}
q_0^\prime (r_\mathrm{v})=0,~~q_0^{\prime \prime} (r_\mathrm{v})>0.
\end{equation} 
When $r_\mathrm{c}=r_\mathrm{v}$ is satisfied, we can obtain the mode frequency from only the background profiles for the marginally stable states due to $\gamma=0$. 
Here, we define the mode frequency for the marginally stable states as 
\begin{equation}
\omega_\mathrm{MS} (m) \equiv m \Omega_\mathrm{0}(r_\mathrm{v}), 
\end{equation}
and $\omega_\mathrm{MS}$ is a real number. 

For the marginally stable states, the effective potential is also a real-number and obtained from only the background profiles as
\begin{eqnarray}
D_\mathrm{MS}(r, m) &\equiv& D(r, \omega_\mathrm{MS}(m), m) \nonumber \\
&=& \frac{B_\mathrm{MS}^\prime}{2} +\frac{B_\mathrm{MS}^2}{4}-C_\mathrm{MS},
\end{eqnarray}
where 
\begin{eqnarray}
B_\mathrm{MS}(r, m) &\equiv& \frac{1}{r}+\frac{F_\mathrm{MS}^\prime}{F_\mathrm{MS}}-\frac{\Omega_0^\prime}{\Omega_0}, \\
C_\mathrm{MS}(r, m) &\equiv& -k_\varphi^2-\frac{\kappa_0^2 -m^2(\Omega_{0, \mathrm{v}}-\Omega_0)^2}{c_0^2}\nonumber \\
&&-\frac{2}{r} \frac{\Omega_0}{(\Omega_{0, \mathrm{v}}-\Omega_0)} \frac{F_\mathrm{MS}^\prime}{F_\mathrm{MS}}, \\
F_\mathrm{MS}(r, m) &\equiv& \frac{\Sigma_0 \Omega_0}{\kappa_0^2 -m^2(\Omega_{0, \mathrm{v}}-\Omega_0)^2}, 
\end{eqnarray}
and the constant $\Omega_0 (r=r_\mathrm{v})$ is denoted by $\Omega_{0, \mathrm{v}}$.
We have used $\omega_\mathrm{MS}=m\Omega_{0, \mathrm{v}}$ in equations (47)-(49) where appropriate. 
Since
\begin{equation}
\frac{F_\mathrm{MS}^\prime(r_\mathrm{v})}{F_\mathrm{MS}(r_\mathrm{v})} =-\frac{q_0^\prime(r_\mathrm{v})}{q_0(r_\mathrm{v})}=0, 
\end{equation}
is satisfied, $D_\mathrm{MS}$ has no singularity at $r=r_\mathrm{v}$. 
This is in consistent with the results and the discussions in \S 5.1. 

\subsection{Search for Parameters of the Marginally Stable States}
We have seen that, for the marginally stable states, it is possible to find the co-rotation radius and the effective potential from only the background vortensity profile. 
We now discuss how we can find the set of the parameters that gives the marginally stable states. 
From equations (46)-(49), $D_\mathrm{MS}$ can be defined even if the system is not exactly in the marginally stable states. 
Figure \ref{fig12} shows $\mathrm{Re}[D]$ and $D_\mathrm{MS}$ for the same case as the left panels in Figure \ref{fig9} ($\gamma /\Omega_\mathrm{n}\simeq 10^{-3}$ for $m=1$). 
$D_\mathrm{MS}$ almost corresponds to $\mathrm{Re}[D]$, but has a smooth profile in the vicinity of the co-rotation radius. 

We derive the parameters for marginally stable states of the system to the RWI without solving equation (26) or (30). 
Since it is expected that the azimuthal mode number is unity for the marginally stable states of the system, we adopt $m=1$ (see Figure \ref{fig7} and \S 4). 
When we integrate 
\begin{equation}
\Xi^{\prime \prime} -D_\mathrm{MS} \Xi =0, 
\end{equation}
$\Xi$ is obtained as a real number function of $r$ because $D_\mathrm{MS}$ is real. 
We fix $h$, $\Gamma$, and $\Delta w$, and look for the value of $\mathcal{A}$ where the solution $\Xi$ satisfies the radiative boundary conditions at the outer and inner radii (see \S 2.4) by using 1D shooting method. 
We denote $\mathcal{A}$ derived in this way by $\mathcal{A}_{\rm MS}$.

\begin{figure*}
\epsscale{2.0}
\begin{center}
\plotone{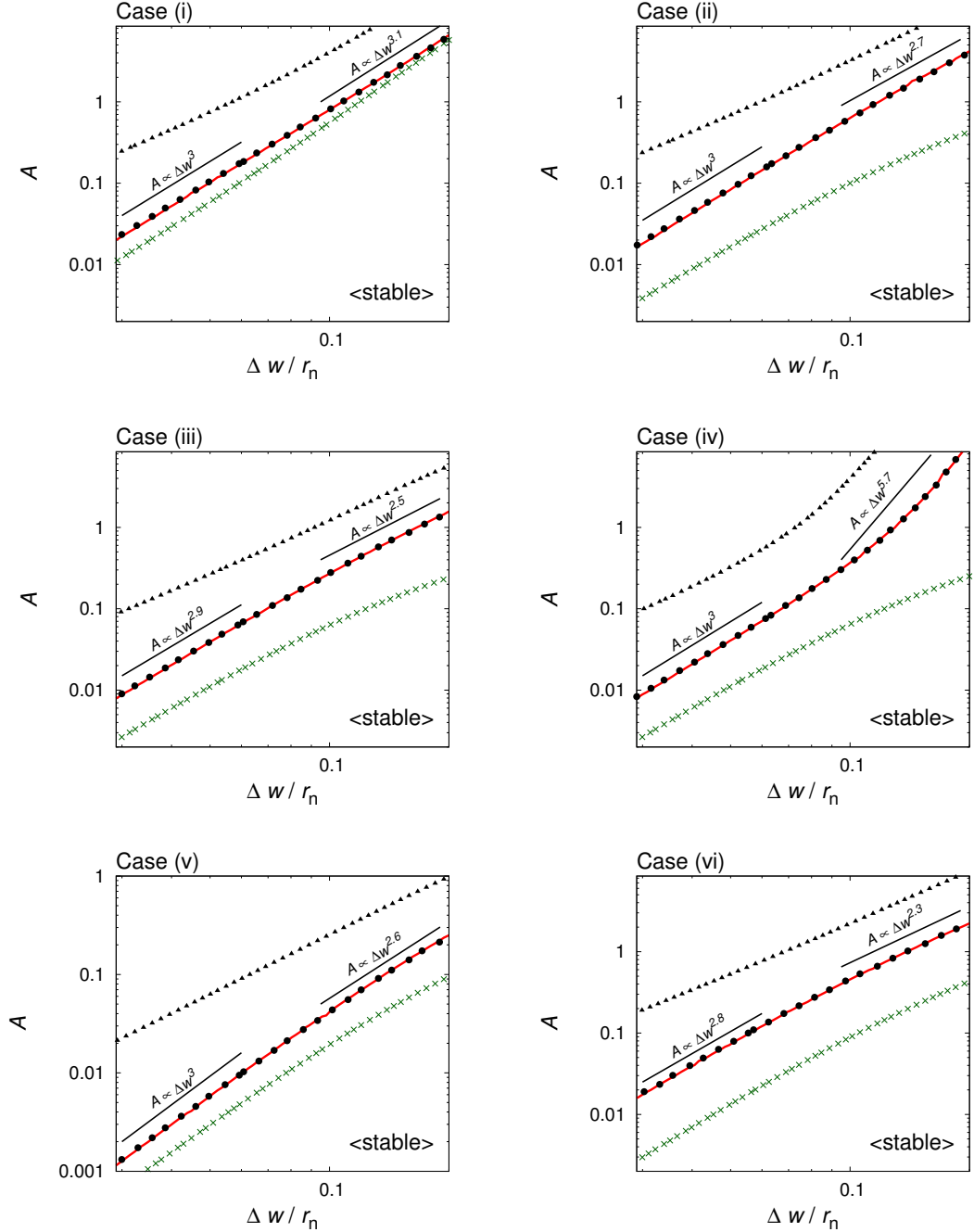}
\vspace{-2cm}
\caption{We show the line with $\gamma_\ast /\Omega_\mathrm{n} = 10^{-3}$ (black dotted line) and the line of $\mathcal{A}_\mathrm{MS}$ by the 1D shooting (red solid line) for each case. 
The line with $\gamma_\ast /\Omega_\mathrm{n} \simeq 10^{-3}$ is fully in accord with the line of $\mathcal{A}_\mathrm{MS}$. 
The lines of dark-green cross and black triangle are same as those in Figure \ref{fig6}. 
For comparison, we show the black solid lines with $\mathcal{A}\propto (\Delta w/ r_\mathrm{n})^\beta$, where $\beta$ is the fitting parameter (see Table 2).  \label{fig13}}
\end{center}
\end{figure*}

By the 1D shooting, we can obtain parameters for the marginally stable states of the system. 
Figure \ref{fig13} shows $\mathcal{A}_{\rm MS}$. 
It is clear that the line of $\mathcal{A}_\mathrm{MS}$ is fully in accord with the line with $\gamma_\ast /\Omega_\mathrm{n}= 10^{-3}$ for all cases. 
The line of $\mathcal{A}_\mathrm{MS}$ can be fitted by 
\begin{equation}
\mathcal{A}_\mathrm{MS}=\alpha \left( \frac{\Delta w}{r_\mathrm{n}} \right)^{\beta} 
\end{equation}
for $0.02\leq \Delta w /r_\mathrm{n}\leq 0.05$ and $0.05\leq \Delta w /r_\mathrm{n}\leq 0.2$. 
We use two ranges of $\Delta w$ for fitting to investigate how the width of the transition of  physical parameters affects the stability against the RWI. 
The fitting parameters, $\alpha$ and $\beta$, for each case are shown in Table 2. 
We obtain $\beta \sim 3$ for all the cases, and in case (v), $\alpha$ is $\sim 8$ times smaller than other cases.  Since $h$ is twice as large as other cases, this suggests that $A_{MS} \propto (\Delta w/h)^3$, especially for small $\Delta w$. 
For large $\Delta w$, $\beta$ is decreased for cases (ii)-(iii) and (v)-(vi), and almost constant for case (i). 
On the other hand, $\beta$ is increased for case (iv). 
Since the case (iv) is for isothermal disks ($\Gamma =1$), these fitting parameters indicate that the effective adiabatic index of the gas $\Gamma$ has a large effect on the stability against the RWI for large $\Delta w$ rather than for small $\Delta w$. 

We note that the shooting method of solving equation (51) can be used to find the parameters $\mathcal{A}$ and $\Delta w$ where the growth rate of the perturbations with a fixed $m$ (not necessarily unity) vanishes. 
In general, when the perturbation with higher $m$ ($m>1)$ is marginally stable, the perturbations with lower $m$ are still unstable. 
Therefore, the stability of the system as a whole is determined by the stability of the modes with $m=1$. 

\subsection{The {\it Necessary And Sufficient} Condition for the Onset of the RWI in Semi-Analytic Form}
Since equation (30) is similar to the stationary Schr\"odinger equation with zero energy, we attempt to analytically derive the condition for the RWI by using the Sommerfeld-Wilson quantization condition \citep{Wilson1915, Sommerfeld1916}. 
The Sommerfeld-Wilson quantization condition is the condition for trapping eigenfunctions in the potential wells. 
Since $\Xi$ with no node is considered to be ground-state, we use the Sommerfeld-Wilson quantization condition for ground states; 
\begin{equation}
\eta (m) \equiv \int^{r_\mathrm{OR}}_{r_\mathrm{IR}} \sqrt{-D_\mathrm{MS}}\, \mathrm{d}r \geq \frac{\pi}{2\sqrt{2}}, 
\end{equation}
where $r_\mathrm{IR}$ is the radius of the inner Rossby turning point and $r_\mathrm{OR}$ is the radius of the outer Rossby turning point. 
The flow is unstable to the RWI when $\eta > \pi/2\sqrt{2}$ and is marginally stable when $\eta = \pi/2\sqrt{2}$. 

\begin{figure*}
\epsscale{2.0}
\plotone{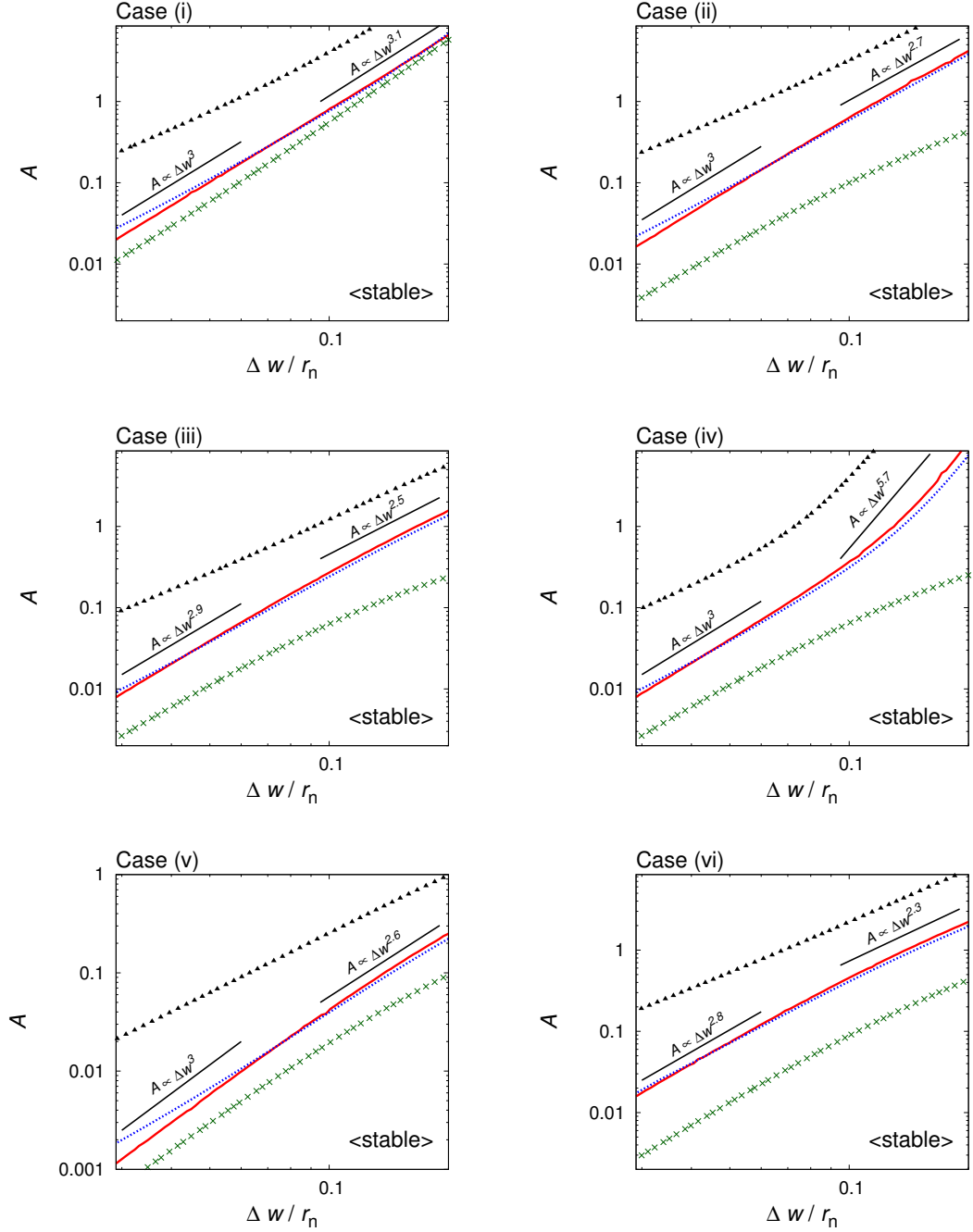}
\vspace{-2cm}
\caption{Similar to Figure \ref{fig13}, but the line with $\eta (m=1) =\pi/(2\sqrt{2})$ is shown as blue dotted line instead of the line with $\gamma_\ast /\Omega_\mathrm{n}= 10^{-3}$. The line with $\eta (m=1) =\pi/(2\sqrt{2})$ and the line of $\mathcal{A}_\mathrm{MS}$ are well fitted except for $\Delta w \ll H$. \label{fig14}}
\end{figure*}

The Sommerfeld-Wilson quantization condition requires the availability of the WKBJ approximation, but it is generally not available for ground states. 
However, the Sommerfeld-Wilson quantization condition is exceptionally available even for ground states when the potential well of the effective potential may be well approximated by a parabolic function with respect to $r$. 
We assume that equation (53) can be used as the condition for the RWI because the potential well of $D_\mathrm{MS}$ looks parabolic as seen in Figure \ref{fig12}. 
We check the the availability of equation (53) for $m=1$ by comparing with the parameters for marginally stable states of the system derived in \S 5.3. 
Figure \ref{fig14} shows the line with $\eta_{m=1} =\pi/(2\sqrt{2})$ for each case in similar manner to Figure \ref{fig13}. 
The lines with $\eta_{m=1} =\pi/(2\sqrt{2})$ are very similar to the lines of $\mathcal{A}_{\rm MS}$ for all cases except for $\Delta w \ll H_0$. 
It is inferred that the potential well of $D_\mathrm{MS}$ deviates from a parabolic potential well for $\Delta w \ll H_0$. 
From the above, equation (53) can be used as the {\it necessary and sufficient} condition for the onset of the RWI in semi-analytic form except for $\Delta w \ll H_0$. 
It is noted that the new condition is also used for other $m$. 
We also note that the new condition is not available for the parameters where the Rayleigh's condition is violated because of the appearance of new singularities in $D_\mathrm{MS}$ (not discussed in detail in this paper). 
The approximation involved in the Sommerfeld-Wilson condition is not satisfied when the rotational instability is expected to occur. 

\begin{table}[t]
\begin{center}
\caption{The fitting parameters by $\mathcal{A}_\mathrm{MS}=\alpha (\Delta w / r_\mathrm{n})^{\beta}$ for each case.\label{tbl1}}
\begin{tabular}{ccccccc}
\tableline\tableline
&&\multicolumn{2}{c}{ $0.02\leq \Delta w /r_\mathrm{n}\leq 0.05$} && \multicolumn{2}{c}{$0.05\leq \Delta w /r_\mathrm{n} \leq 0.2$}\\
\tableline
Case &&$\alpha$&$\beta$& &$\alpha$ &$\beta$ \\
\tableline
i &&8.08 $\times 10^2$ &3.00 &&$9.07 \times 10^2$ &3.06 \\
ii &&6.95 $\times 10^2$&3.01 &&$3.33 \times 10^2$ &2.72 \\
iii &&2.58 $\times 10^2$&2.93 &&$9.32 \times 10$ &2.53 \\
iv &&3.25 $\times 10^2$&3.00 &&$1.06 \times 10^5$ &5.72 \\
v &&4.00 $\times 10$&2.95 &&$1.63 \times 10$ &2.58 \\
vi && 3.29 $\times 10^2$ &2.80 &&$9.31 \times 10$ &2.31 \\
\tableline
\end{tabular}
\end{center}
\end{table}

\subsection{Consideration of Other Physical Effects}
In this paper, we have made several simplifying assumptions. 
We assume barotropic disks with the flat base profile $\mathfrak{f}=0$. 
We also neglect the effects of viscosity and self-gravity of the disk. 
Here, we briefly discuss how these assumptions affect the stability and present the prospects for the applications to realistic systems. 

To investigate the effects of different base profiles, we have checked the marginal stability conditions for the cases with $f=1.5$, as in the case of the minimum mass solar nebula \citep{1981PThPS..70...35H}. 
We have found that the results for $\mathfrak{f}=1.5$ are almost same as the results for $\mathfrak{f}=0$. 
This indicates that the global radial variations of surface mass density have only minor effects on the onset of the RWI, confirming that rapid radial variations of the background physical parameters (such as bump, gap, step jump) are essential. 
We therefore expect that the new condition can be used as the {\it necessary and sufficient} condition for the onset of the RWI even if the base profile is not flat ($\mathfrak{f}\neq 0$). 

Non-barotropic effects on the RWI have been discussed in several studies \citep{1999ApJ...513..805L, 2000ApJ...533.1023L, 2013ApJ...765...84L}. 
Those previous works show that the RWI is more likely to be unstable in non-barotropic flow, probably due to the baroclinic term acting as the source of vortensity \citep{1999ApJ...513..805L}. 
The growth rate of the RWI for non-barotropic flow is slightly larger than that for barotropic flow \citep{2000ApJ...533.1023L, 2013ApJ...765...84L}. 
Moreover, $\mathcal{A}_\mathrm{MS}$ for non-barotropic flow is slightly smaller (see the Figure 9 of \citealt{2000ApJ...533.1023L}). 
It seems, however, that the difference of $\mathcal{A}_\mathrm{MS}$ between the barotropic and non-barotropic cases is small. 
We therefore expect that our new condition can be reasonably applicable even for non-barotropic cases. 
Quantitative predictions for the stability require full explorations of parameter spaces and mode numbers. 

In this work, we have neglected the effects of viscosity and self-gravity. 
\citet{2014MNRAS.437..575L} shows that the RWI is largely unaffected by viscosity and occurs in the linear regime. 
The effect of self-gravity is discussed by some studies \citep{2013MNRAS.429..529L, 2015arXiv150308470Y}. 
\citet{2013MNRAS.429..529L} shows that the onset of the RWI is depressed for $m\mathcal{Q}H_0/r_\mathrm{n}\lesssim \pi/2$, where $\mathcal{Q}$ is the Toomre $\mathcal{Q}$ parameter of the disks \citep{1964ApJ...139.1217T}. 
In other words, the RWI does not occur in heavy disks. 

In applications to realistic systems and observed protoplanetary disks, it is necessary to consider various "base" models, non-barotropy, viscosity, self-gravity and a number of other physical effects. 
It seems that the slope of the background profiles, non-barotropy and viscosity do not seriously affect our results. 
Considering the stabilizing effect of self-gravity, our condition may act as a necessary condition for the RWI even if self-gravity of the disks is large. 
Detailed investigations considering other physical effects in both linear and non-linear regimes are necessary to understand the RWI in realistic systems. 

\subsection{Relation of the RWI with Associated Known Conditions}
\begin{figure}
\epsscale{1.0}
\plotone{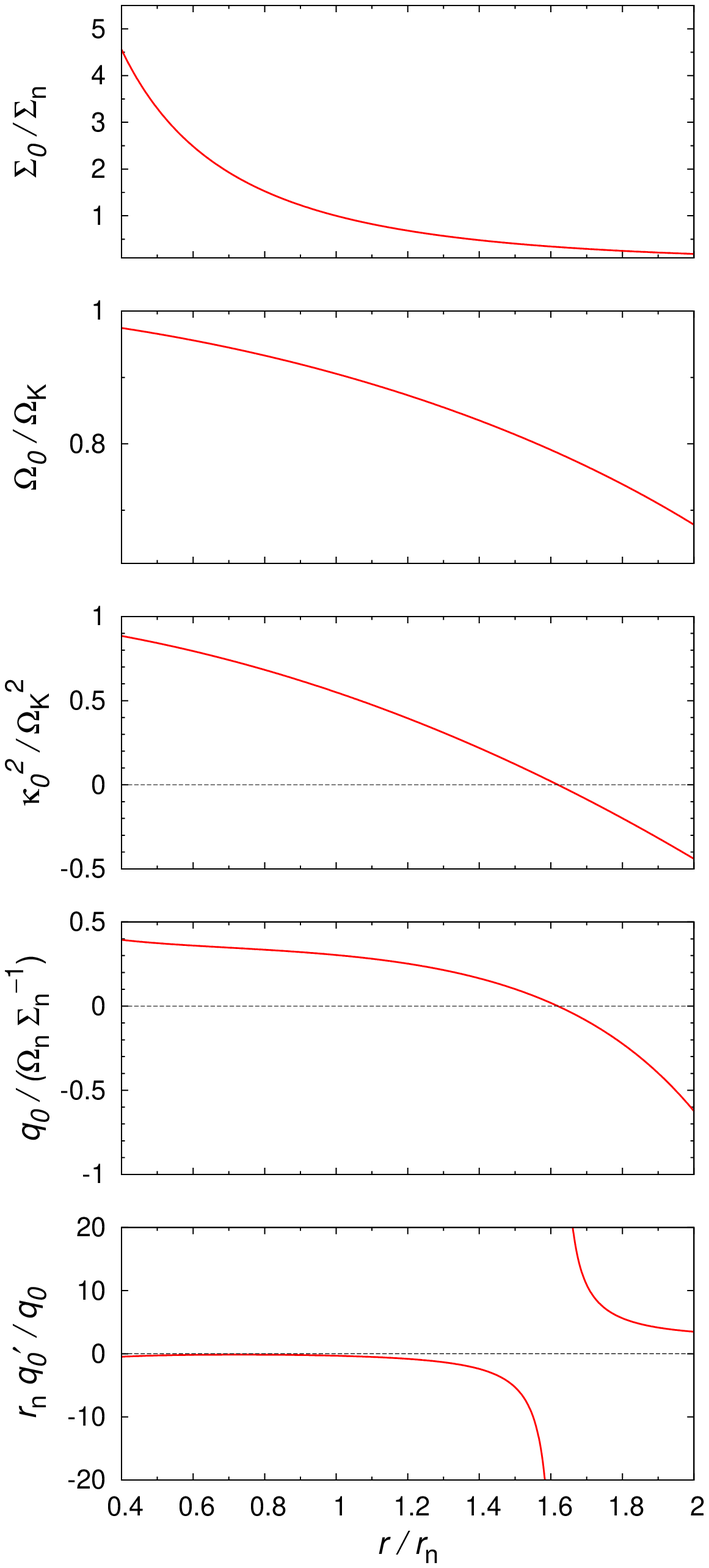}
\caption{Similar to Figure \ref{fig1}, but for the similarity solution with $h=0.3$ and $\Gamma=1.0$. 
This profile violate the Rayleigh's condition, while has no $q_0$ minimum. \label{fig15}}
\end{figure}

From numerical results (\S 4) and analytic considerations (\S 5), it is now clear that the Lovelace's condition is the necessary condition for the onset of the RWI. 
Within the parameter space we have explored, it may also be possible to say that the Rayleigh's condition is the sufficient condition for the RWI. 
However, we actually find an example where the Rayleigh's condition is violated but the system is stable to the RWI. 
This is the case for the similarity solution of the accretion disk \citep{1974MNRAS.168..603L}; 
\begin{equation}
\frac{\Sigma_0}{\Sigma_\mathrm{n}}=\frac{1}{r}\exp \left[ -\left(\frac{r-r_\mathrm{n}}{r_\mathrm{n}} \right) \right] 
\end{equation}
with $h=0.3$ and $\Gamma=1$. 
The Rayleigh's condition is violated in this case \citep{2014ApJ...787...37O}. 
Figure \ref{fig15} shows the background profiles of this similarity solution in a similar manner to Figure \ref{fig1}. 
Due to the absence of the $q_0$ minimum, this similarity solution does not satisfy the Lovelace's condition, and is stable to the RWI. 
Therefore, the Rayleigh's condition is not the sufficient condition for the onset of the RWI. 

\section{Physical Nature of the RWI} 
In this section, we attempt to explore the physical nature of the RWI based on the eigenfunction obtained from our linear stability analyses.
One way to explain the physics of the RWI is to use a wave action, which is the adiabatic invariant related to angular momentum of the perturbations. 
\citet{1987MNRAS.228....1N} discusses the Drury-Papaloizou-Pringle instability under the shearing sheet approximation, and \citet{2008MNRAS.387..446T} does the RWI under the WKBJ approximation. 
Here, we attempt to understand the physics of the RWI from the slightly different point of view. 

\subsection{Perturbation patterns in the Real Space}
\begin{figure*}
\epsscale{2.0}
\plotone{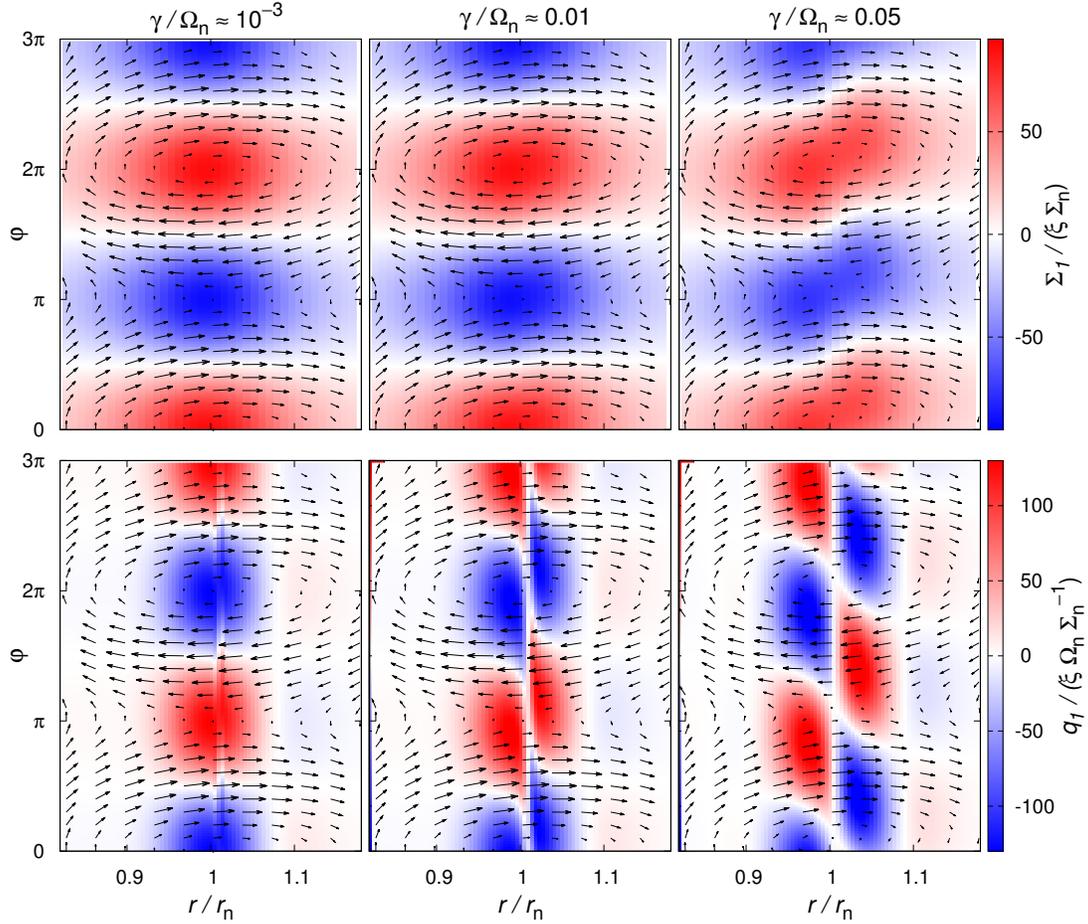}
\caption{$\Sigma_1$ (top panels) and $q_1$ (bottom panels) are shown in the $r$-$\varphi$ plane with $\gamma /\Omega_\mathrm{K}\simeq 10^{-3}$ ($\mathcal{A}=0.04$; left panels), $0.01$ ($\mathcal{A}=0.05$; middle panels) and $0.05$ ($\mathcal{A}=0.1233$; right panels) for $m=1$ in the rotating frame with the co-rotation radius for case (iii) with $\Delta w=0.05 r_\mathrm{n}$. 
Black arrows show velocity field ${\bf v_0} +\xi {\bf v_1}$ and we adopt $\xi=0.025$. 
For legibility, $\varphi$ is shown in the range of $0\leq \varphi \leq 3\pi$. 
\label{fig16}}
\end{figure*}

In the real space, surface mass density perturbation and vortensity perturbation are given by $\Sigma_1 (r, \varphi) = \mathrm{Re}[\Sigma_1(r) \exp(im\varphi)]$ and $q_1 (r, \varphi) = \mathrm{Re}[q_1(r) \exp(im\varphi)]$ respectively. 
Figure \ref{fig16} shows $\Sigma_1 (r, \varphi)$ (top panels) and $q_1 (r, \varphi)$ (bottom panels) for $\gamma /\Omega_\mathrm{K}=10^{-3}$ ($\mathcal{A}=0.04$; left panels), $0.01$ ($\mathcal{A}=0.05$; middle panels) and $0.05$ ($\mathcal{A}=0.1233$; right panels) for case (iii) with $\Delta w=0.05r_\mathrm{n}$ and $m=1$ in the frame that rigidly rotates at the rotation velocity of the co-rotation radius. 
We also note that the normalization constant of the perturbation $\xi$ is given by 0.025 (see Section 5.1 for the definition of $\xi$) in Figure \ref{fig16}. 
Since $r_\mathrm{c}\simeq r_\mathrm{n}$ is satisfied in those cases, the background shear velocity is in the $-\varphi$ direction for $r \gtrsim r_\mathrm{n}$ and in the $+\varphi$ direction for $r\lesssim r_\mathrm{n}$. 
There are two perturbation patterns of both $\Sigma_1$ and $q_1$ in the $\varphi$ direction, and it appears that the patterns hold the co-rotation radius in between. 
The patterns of the perturbations are stationary in the rotating frame. 
It should be noted that $\varphi$ is shown in the range of $0\leq \varphi \leq 3\pi$ for legibility. 

First, we discuss the $\Sigma_1$ patterns. 
Positive $\Sigma_1$ regions have clockwise velocity field around themselves, while negative $\Sigma_1$ regions have anti-clockwise velocity field. 
It appears that the $\Sigma_1$ patterns get distorted due to the background shear, and the deformation gets larger with the growth rate. 
Due to the deformation, there is more incoming flow at the positive $\Sigma_1$ regions than outgoing flow. 
On the other hand, there is more outgoing flow at the negative $\Sigma_1$ regions than incoming flow. 
This certainly describes the unstable states. 

Since vortensity is the key physical quantity of the RWI as seen \S 5, we discuss from the point of the view of $q_1$ rather than $\Sigma_1$. 
Positive $q_1$ regions roughly corresponds to the negative $\Sigma_1$ regions, and vice versa. 
In other words, $q_1$ has an anti-correlation with $\Sigma_1$. 
It is possible to see that there are two chains of the vortensity perturbation in the $\varphi$ direction, located side by side around the co-rotation radius. 
The two $q_1$ patterns exhibit the difference in phase in the $\varphi$ direction, and the phase difference increases as the growth rate increases. 
The first-order form of equation (4), which express the conservation of vortensity with respect to fluid elements, is 
\begin{equation}
\henb{q_1}{t} +v_{r1}\henb{q_0}{r} +\Omega_0 \henb{q_1}{\varphi}  =0.
\end{equation}
The second term of equation (55) can be interpreted as the advection of the background vortensity in the radial direction by the perturbed velocity field, while the third term represents the advection of the vortensity perturbation in the azimuthal direction due to the background flow. 
From bottom panels in Figure \ref{fig16}, it is possible to observe that, in the regions where the radial velocity perturbation is positive (e.g., $\varphi \sim \pi/2$) and $r/r_n>1$ ($\sim$ outside the corotation radius), the vortensity perturbation is negative. 
Since the background vortensity is minimum at $r \sim r_\mathrm{n}$, this velocity perturbation will further decrease the vortensity perturbation and therefore the perturbation grows. 
In the region where the radial velocity perturbation is negative ($\varphi \sim 3\pi/2$) and $r/r_n<1$ (inside the co-rotation radius), the vortensity perturbation is also negative and the perturbation also grows. 
The phase difference between the vortensity perturbations at $r<r_\mathrm{c}$ and $r>r_\mathrm{c}$ is essentially important for the RWI to occur.

\subsection{Analogy of the Shear Instability in Incompressible Flow}

It is verified that the Lovelace's condition is the necessary condition for the RWI (see \S 5.6). 
The Lovelace's condition is considered to be identical as the Rayleigh's inflection-point theorem \citep{Rayleigh1880}, which is the necessary condition for the shear instability in incompressible flow. 
Therefore, the RWI is likely to be one of the shear instabilities. 
It is known that the shear instabilities in incompressible flow is interpreted by the wave interaction \citep{Drazin1981, Hayashi1987}. 
Due to the similarities between incompressible flow in shallow water and barotropic flow, the physical nature of the RWI is interpreted from the point of view of the wave interaction \citep{2010A&A...521A..25U}. 

\begin{figure}
\epsscale{1.2}
\plotone{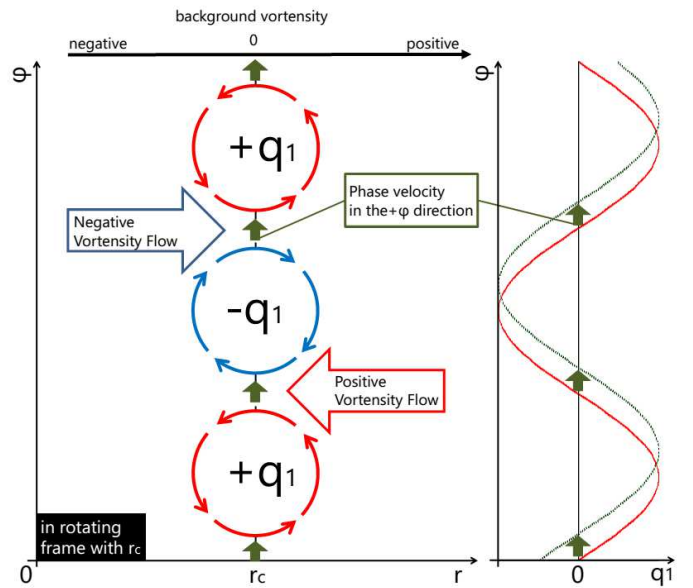}
\caption{The schematic figure of the Rossby wave. \label{fig17}}
\end{figure}

The $q_1$ patterns are considered to be identical as the ''Rossby wave'' in geophysics \citep{Rossby1939}. 
The Rossby wave is a neutral wave and exists when the background vortensity profile has a gradient. 
Hereafter, we call the $q_1$ pattern the Rossby wave. 
In order to introduce the Rossby wave, we first consider the system which satisfies $q_0^\prime >0$ and $q_0(r_\mathrm{c})=0$, being in the rotating frame with the co-rotation radius. 
In this system, there is one Rossby wave and velocity perturbations induced by the Rossby wave as Figure \ref{fig17}. 
Figure \ref{fig17} is the schematic figure of the Rossby wave. 
Since vortensity is conserved with respect to fluid elements and $q_0^\prime >0$ is satisfied, incoming flows to the co-rotation radius from the outside have positive vortensity and those from the inside have negative vortensity. 
These flows induce phase velocity of the Rossby wave in the $+\varphi$ direction.
On the other hand, the phase velocity is in the $-\varphi$ direction for $q_0^\prime <0$. 
The speed of the phase velocity depends on $q_0^\prime$ at the radius of the Rossby wave. 

We next explain the physical nature of the RWI using two Rossby waves. 
When $q_0$ has an extremum and the co-rotation radius located at the $q_0$ extremum, there are two Rossby waves at the radii between which the extremum of $q_0$. 
We fix $\Delta w$ and vary $\mathcal{A}$. 
From Figure \ref{fig16}, the pattern of the two Rossby waves is required to be stationary in the rotating frame with the co-rotation radius for the unstable states. 
We define that $v_\mathrm{phase}$ is the phase speed of the Rossby waves and $v_\mathrm{shear}$ is the background shear speed. 
Since $v_\mathrm{phase}$ depends on $q_0^\prime$ at the radius of the Rossby waves, $v_\mathrm{phase}$ increases and $v_\mathrm{shear}$ decreases with the increase of $\mathcal{A}$ basically. 

\begin{figure*}
\epsscale{2.0}
\plotone{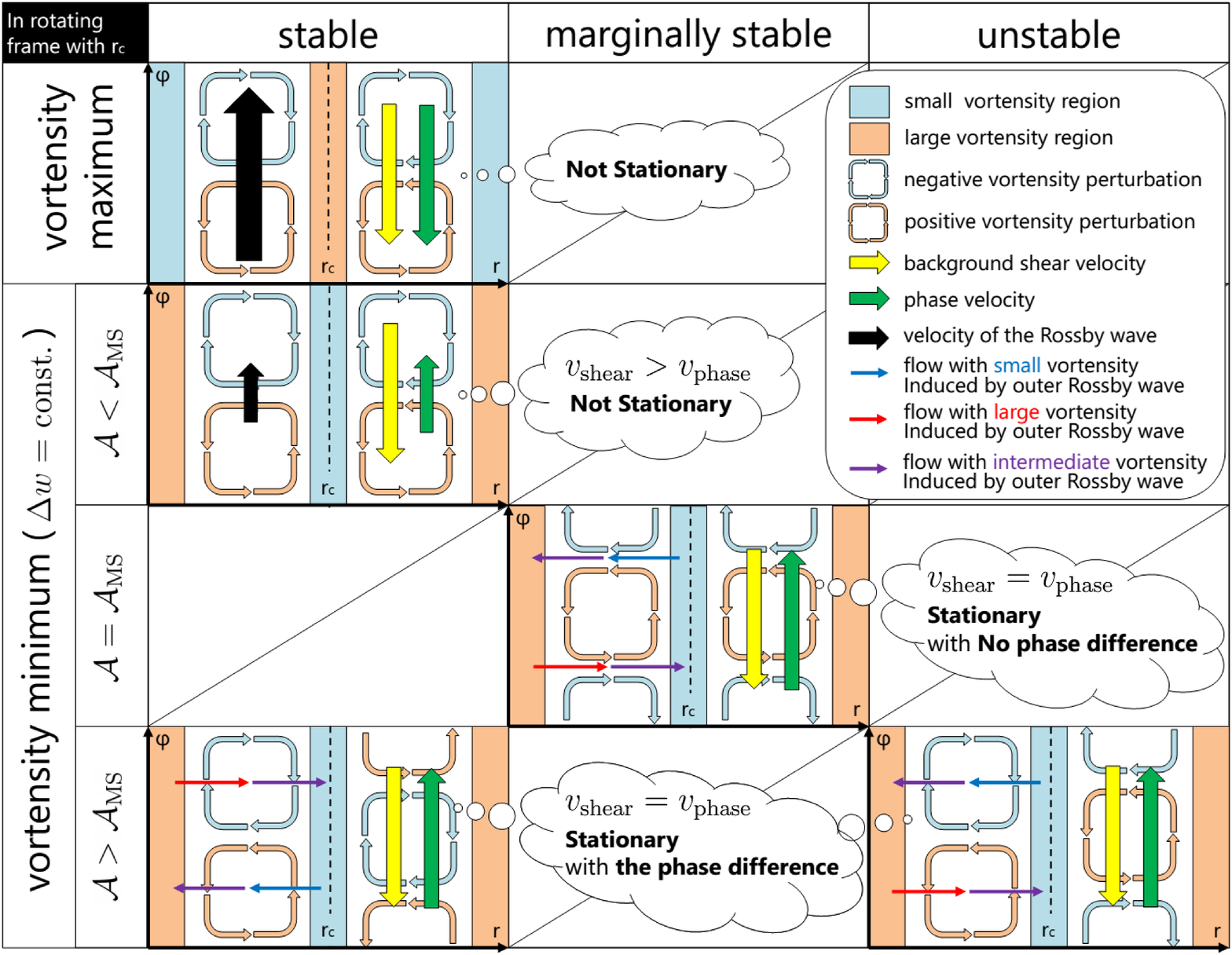}
\caption{The schematic picture of the physical nature of the RWI. \label{fig18}}
\end{figure*}

Figure \ref{fig18} shows the schematic picture of the RWI for each case. 
For a $q_0$ maximum (first row of Figure \ref{fig18}), the phase velocity is in the same direction of the background shear velocity. 
The position of the two Rossby waves cannot be stationary. 
Therefore, it is impossible that the co-rotation radius is located in the vicinity of the $q_0$ maximum for the unstable states. 

For a $q_0$ minimum (from second to fourth rows of Figure \ref{fig18}), the phase velocity is in the opposite direction of the background shear velocity. 
For $\mathcal{A}<\mathcal{A}_\mathrm{MS}$ (second row of Figure \ref{fig18}), $v_\mathrm{phase}$ is less than $v_\mathrm{shear}$. 
The position of the two Rossby waves is still not stationary, and the system is stable to the RWI. 

For $\mathcal{A}=\mathcal{A}_\mathrm{MS}$ (third row of Figure \ref{fig18}), $v_\mathrm{phase}$ is equal to $v_\mathrm{shear}$, and the position of the two Rossby waves is stationary. 
In this case, there is no phase difference between the Rossby waves. 
We take notice of the inner Rossby wave and the flows induced by the outer Rossby wave. 
Due to no phase difference, the net flux of the vortensity into the positive or negative $q_1$ regions is equal to zero. 
The system is marginally stable to the RWI. 

For $\mathcal{A}>\mathcal{A}_\mathrm{MS}$ (fourth row of Figure \ref{fig18}), $v_\mathrm{phase}$ is equal to $v_\mathrm{shear}$ but the phases of the two Rossby waves are different. 
The phase velocity of the Rossby waves does not seem to increase even if we increase the strength of the bump probably because the interaction between the two Rossby waves is strong.
The position of the two Rossby waves is stationary with the non-zero phase difference. 
We can have two kinds of the phase differences. 
One is the phase difference that makes the RWI unstable, and the other that makes the RWI stable. 
We take notice of the inner Rossby wave and the flows induced by the outer Rossby wave. 
When the outer Rossby wave has a larger phase than the inner Rossby wave (the positive phase difference), the net flux of the vortensity into the positive $q_1$ regions is negative, and vice versa. 
In that case, the system is stable to the RWI. 
On the other hand, when the outer Rossby wave has a smaller phase than the inner Rossby wave (the negative phase difference), the net flux of the vortensity into the positive $q_1$ regions is positive, and vice versa. 
In that case, the system is unstable to the RWI. 

These explanations and Figure \ref{fig18} are consistent with \citet{2010A&A...521A..25U} and our results of the linear stability analyses. 
In addition, since $v_\mathrm{phase}$ depends on $q_0^\prime$ at the radius of the Rossby waves and $q_0^\prime =0$ is satisfied at the co-rotation radius, $q_0^{\prime \prime}(r_\mathrm{v})$ should be large enough to achieve $v_\mathrm{phase}=v_\mathrm{shear}$. 
Those are why the co-rotation radius is located at the background vortensity minimum with large concavity for the marginally stable states. 

\section{Conclusions}
We perform the linear stability analyses for representative 6 types of the background flow on a wide parameter space, and examine the marginally stable condition for the RWI and the physical nature of the RWI.
We find that the co-rotation radius is located at the background vortensity minimum with large concavity for marginally stable states to the RWI. 
Using these results, the stability against the RWI can be checked by at least the 1D-shooting. 
We newly derive the {\it necessary and sufficient} condition for the onset of the RWI in semi-analytic form on the analogy of the quantum mechanics. 
The new condition is considered to be available except for $\Delta w \ll H_0$ in general. 
It is verified that the Lovelace's condition is the necessary condition for the RWI. 
On the other hand, we prove that the Rayleigh's condition is not the sufficient condition for the RWI. 
We expect that the new condition will help to know the stability of the observed protoplanetary disks against the RWI. 
Applications of the new condition to realistic systems require more detailed investigations considering other physical effects in both linear and non-linear regimes. 

\acknowledgments
We gratefully acknowledge Shinichi Takehiro, Michio Yamada and Wladimir Lyra for their detailed comments. 
We are also grateful to the referee who helped improve the quality of the manuscript.
This work was partially supported by JSPS KAKENHI Grant Numbers 15J01554 (T.O.), 26800106, 23103004, 15H02074 (T.M.), 26103704 (T.T.), 23103005 and 25400229 (H.N.). 
A part of numerical computation in this work was carried out at the Yukawa Institute Computer Facility.






\appendix
\section {Method for Solving the Linear Stability Analysis}
We show our method to solve equation (26). 
We discretize equation (26) on a grid $\mathfrak{i}= 0, ..., N$ over equal intervals $\Delta r$, and define $r_\mathfrak{i}$ as the radius at the $\mathfrak{i}$-th grid. 
Equation (26) is rewritten as 
\begin{equation}
\sum_{\mathfrak{j}=0}^N W_{\mathfrak{i}, \mathfrak{j}}(\omega)\Psi (r_\mathfrak{j}) = 0,
\end{equation}
where ${\bf W}$ is the $(N+1) \times (N+1)$ matrix and $W_{\mathfrak{i}, \mathfrak{j}}$ denotes the element of ${\bf W}$ in the $\mathfrak{i}$-th row and the $\mathfrak{j}$-th column. 
When we discretize equation (26) as 
\begin{eqnarray}
\left. \Psi^{\prime \prime} \right|_{r=r_\mathfrak{i}} \rightarrow \frac{\Psi(r_{\mathfrak{i}+1})-2\Psi(r_{\mathfrak{i}})+\Psi(r_{\mathfrak{i}-1})}{(\Delta r)^2} \nonumber
\end{eqnarray}
and 
\begin{eqnarray}
&\left. \Psi^\prime \right|_{r=r_\mathfrak{i}} \rightarrow \frac{\Psi(r_{\mathfrak{i}+1})-\Psi(r_{\mathfrak{i}-1})}{2\Delta r} ~\mathrm{(without~boundaries)},  \nonumber \\
&\left. \Psi^\prime \right|_{r=r_1} \rightarrow \frac{\Psi(r_{1})-\Psi(r_{0})}{\Delta r} ~\mathrm{(at~ the~inner~boundary)},  \nonumber \\
&\left. \Psi^\prime \right|_{r=r_{N-1}} \rightarrow \frac{\Psi(r_{N})-\Psi(r_{N-1})}{\Delta r} ~\mathrm{(at~ the~outer~boundary)},  
\end{eqnarray}
${\bf W}$ is given as a tridiagonal matrix; $W_{\mathfrak{i}, \mathfrak{j}}=0$ for $|\mathfrak{i}-\mathfrak{j}|>1$. 

We can obtain the elements of $W$ in the $0$-th and $N$-th rows from boundary conditions. 
The elements of $W$ in the $0$-th row are
\begin{eqnarray}
W_{0, 0} =1,~~W_{0, 1}=\Delta r \frac{-B(r_1)+ i \sqrt{4C(r_1)-B(r_1)^2}}{2}-1
\end{eqnarray}
for $m\neq 1$, and 
\begin{eqnarray}
W_{0, 0} =1,~~W_{0, 1}=\frac{\Delta r}{2} \left\{-\left[B(r_1)-\frac{1}{r_1} \right] +\sqrt{\left[B(r_1)-\frac{1}{r_1} \right]^2 -4C(r_1)} \right\} -1 
\end{eqnarray}
for $m=1$.
The elements of $W$ in the $N$-th row are 
\begin{eqnarray}
W_{N, N-1} =\Delta r \frac{B(r_{N-1})- i \sqrt{4C(r_{N-1})-B(r_{N-1})^2}}{2}-1,~~W_{N, N}=1. 
\end{eqnarray}
The elements of $W$ in the other columns are obtained from equation (26) as
\begin{eqnarray}
&W_{\mathfrak{i}, \mathfrak{i}-1}=1-\frac{\Delta r}{2}B(r_\mathfrak{i}) \nonumber \\
&W_{\mathfrak{i}, \mathfrak{i}}=(\Delta r)^2\, C(r_\mathfrak{i}) -2 \nonumber \\
&W_{\mathfrak{i}, \mathfrak{i}+1}=1+\frac{\Delta r}{2}B(r_\mathfrak{i}) 
\end{eqnarray}
for $1\leq \mathfrak{i} \leq N-1$. 

Our purpose is to obtain the eigenvalue for the unstable mode frequency $\omega_\mathrm{s}$ ($\mathrm{Im}[\omega_\mathrm{s}]>0$) and the eigenfunction $\Psi_\mathrm{s}(r)$ for $\omega=\omega_\mathrm{s}$. 
For $\omega=\omega_\mathrm{s}$, $|\mathrm{det}[{\bf W}(\omega_\mathrm{s})]|=0$ is satisfied. 
Because the matrix ${\bf W}(\omega)$ is non-linear of $\omega$ and its elements are complex numbers, the problem is a non-linear complex eigenvalue problem. 
Here, we call $\omega_\mathrm{s}$ the unstable root. 
We have two steps to solve this problem. 

First, we evaluate the unstable root. 
We define the sequence of the numbers \mbox{\{ $u_i$ \}}, where 
\begin{eqnarray}
u_0&=&1, ~~
u_1=W_{0, 0}=1 \nonumber \\
u_{\mathfrak{i}+2}&=&W_{\mathfrak{i}+1, \mathfrak{i}+2}\, u_{\mathfrak{i}+1}-W_{\mathfrak{i}, \mathfrak{i}+1}W_{\mathfrak{i}+1, \mathfrak{i}}\, u_\mathfrak{i}.
\end{eqnarray}
Since ${\bf W}$ is a tridiagonal matrix, the determinant of ${\bf W}$ is 
\begin{equation}
\mathrm{det}[W(\omega)]=u_{N+1}.
\end{equation}
We estimate roughly the unstable root. 
Thereafter, we locate the unstable root exactly by using the Muller's method, which is the one of the root-finding algorithms and available even if we have a complex function or are ignorant of the formula of the function. 

\begin{figure}
\begin{center}
\epsscale{0.8}
\plotone{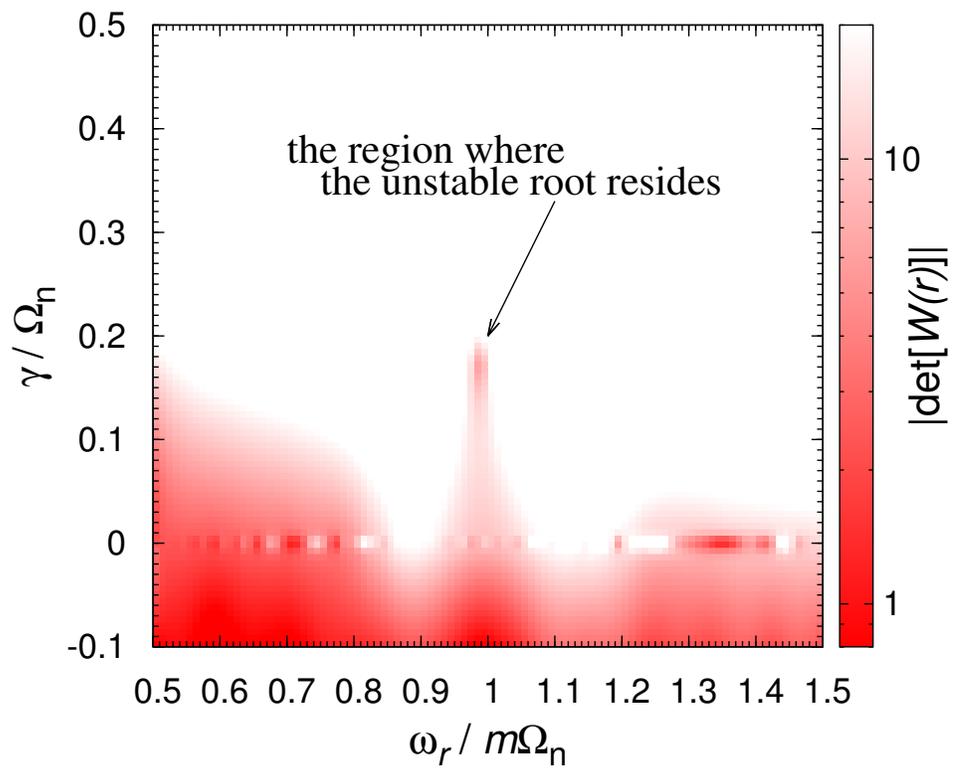}
\vspace{2cm}
\caption{The distribution of $|\mathrm{det}[W(r)]|$ in the complex plane for the same case as Figure \ref{fig3}. 
We can find the region where the unstable root resides around $\omega_\mathrm{s} /\Omega_\mathrm{n}=(m, 0.2)$. 
The accurate unstable root is derived by using the Muller's method is located at $\omega_\mathrm{s}/\Omega_\mathrm{n}=(0.986791 \times m, 0.172961)$. \label{A1}}
\end{center}
\end{figure}

Figure \ref{A1} shows $|\mathrm{det}[{\bf W}(\omega)]|$ in the complex plane for the same case as Figure \ref{fig3} and $m=3$. 
We can find the region where the unstable root resides around $\omega_\mathrm{s}/\Omega_\mathrm{n}=(1.0 \times 3, 0.2)$. 
We obtain $\omega_\mathrm{s}/\Omega_\mathrm{n}=(0.986791 \times 3, 0.172961)$ by the Muller's method. 

Second, we calculate the eigenfunction $\Psi_s(r)$. 
The problem now becomes a linear complex eigenvalue problem; 
\begin{eqnarray}
\left\{
\begin{array}{l}
\sum_{\mathfrak{j}=0}^{N} W_{\mathfrak{i}, \mathfrak{j}}(\omega_\mathrm{s})\Psi(r_\mathfrak{j})=\epsilon \Psi(r_\mathfrak{i}),  \\
\epsilon =0.
\end{array}
\right.
\end{eqnarray}
We solve equation (A9) by using the LAPACK, and obtain $\Psi_s(r)$. 
We have checked that our results derived from these methods are in agreement with the results in \citet{2000ApJ...533.1023L, 2001ApJ...551..874L} for some parameter sets. 

We have to refer that it is difficult to obtain the unstable root for $\mathrm{Re}[\omega_\mathrm{s}]/\Omega_\mathrm{n}\ll 1$ with the above methods. 
Actually, there are enormous neutral roots of $|\mathrm{det}[{\bf W}(\omega)]|$ with $\gamma =0$. 
For $\mathrm{Re}[\omega_\mathrm{s}]/\Omega_\mathrm{n}\lesssim 10^{-5}$, the unstable root is buried in enormous neutral roots. 
In spite of this difficulty, we can certainly distinguish the unstable root from the neutral roots for $\mathrm{Re}[\omega_\mathrm{s}]/\Omega_\mathrm{n}\geq 10^{-4}$.




\clearpage



\clearpage

\end{document}